\documentclass[
draft    
,numberedheadings 
,bibliocites
  ]
  {aipproc}

\layoutstyle{6x9}

\usepackage{amssymb}
\usepackage{eucal}
\usepackage{mathrsfs}
\usepackage{wrapfig}

\newcommand{\scrif}{{{\mathscr I}^{+}}}
\newcommand{\scrip}{{{\mathscr I}^{-}}}
\newcommand{\B}{{\mathscr B}} 
\newcommand{\A}{{\mathcal A}}
\newcommand{\HH}{{\mathcal H}}
\newcommand{\Hf}{{{\mathscr H}^{+}}} 
\newcommand{\Lie}{{\mathcal L}}

\newcommand{\R}{{\mathbb R}}

\newcommand{\MM}{{\mathscr M}}

\newcommand{\as}{{\acute s}}
\newcommand{\vv}{{v}}
\newcommand{\uu}{{u}}

\begin{document}

\title{Black Holes Reconsidered}

\classification{04.70.Bw,
04.70.Dy,
04.62.+v}

\keywords{black holes, problem of teleology, quantum fields in curved space--time, Hawking radiation}

\author{Adam D. Helfer}{
  address={Department of Mathematics, University of Missouri, Columbia, MO 65211, U.S.A.}
}

\begin{abstract}
I reexamine some elements of the theory of black holes, pointing out where the usual treatment seems to be adequate and where there is reason to want to improve it.  Some of the problematic elements of the theory can be clarified by studying holonomies relating the neighborhood of the horizon to the regime occupied by distant observers.
\end{abstract}

\maketitle

\section*{Introduction}

The current basis for the theoretical investigation of black holes largely grew out of the tremendously productive period 1963--1974, sometimes called the Golden Age of Black-Hole Theory, during which key concepts and results were obtained; from these, an extraordinarily rich and exciting
field has developed.

There have remained, however, several awkward elements of the foundations:  the {\em problem of teleology} (that the strict definition of a  black hole requires the knowledge of the entire
 space--time), difficulties fleshing out the notion of black-hole thermodynamics, and problems surrounding Hawking's prediction of black-hole radiation.  For many investigations, these concerns are not directly relevant, but for others they are.  
The problem of teleology, for instance, underlies the difficulties in formulating an observational criterion for the identification of black holes, figures in many arguments involving black holes and causal processes, and is ultimately responsible for the
attempts to find useful substitutes for the event horizon (trapping, isolated and dynamical horizons).
And the ``information paradox'' and ``holography'' are evidently bound up with Hawking's argument for black-hole radiation. 

In this paper I take up these awkard elements:  partly, my aim is to review the usual development of the theory, pointing out open questions and concerns; partly, it is to describe some new ideas which help either to resolve or to elucidate the difficulties.  In particular, perhaps surprisingly, studying the holonomy around paths linking the neighborhood of the horizon to the regime distant observers occupy both brings out important features of the incipient black hole in a non-teleological way, and clarifies the trans-Planckian problem associated with Hawking's prediction.  It also helps approach the problem of delineating just where the classical treatment of black holes becomes inadequate.

While at present the quantum issues appear to be almost 
beyond hope of observation, the case for
reviewing the structure of the neighborhood of the event horizon and casting it in terms which are at least in principle observable rather than teleological is more pressing.  There is a real prospect that within the next decade or so we will begin to be able to resolve some of the structure of the (presumed) supermassive black hole at the center of the Milky Way with submillimeter VLBI networks~\citep{Doeleman:2009te}.

Section 1 reviews black holes and causal structure, emphasizing the conceptual issues.

Section 2 introduces the theory of the holonomy around paths relating the neighborhood of the horizon to the regime occupied by distant observers.  It is shown that there are certain {\em universal} features, which should be observable, associated with the development of an incipient event horizon.

Section 3 is a sketch of the elements of quantum field theory in curved space--time, without taking up the question of what special effects might be due to black holes.

Section 4 reviews Hawking's prediction of thermal radiation from black holes, with attention to the problematic elements of the derivation, and also to some indirect arguments which have been suggested for it.

Section 5 probes deeper into the relation between black holes and quantum theory.  It is first shown that it is reasonable to expect that the close neighborhood of an event horizon will not have an operational interpretation as a classical portion of space--time.  Then it is shown that the trans-Planckian problem is promoted from a virtual to a real one if we consider any of three common quantum effects:  the results of sequences of measurements; the nonlinear interactions of quantum fields; and Casimir energies.  These points lead me to suggest that the current theory of quantum fields in curved space--time, although so natural as to appear unexceptionable, may need to be reconsidered.

The remainder of this section is devoted to a brief discussion of the problem of teleology and associated terminological issues.  

The conventions used here are those of \citet{Penrose:1986ca}; the metric has signature $+{}-{}-{}-$.  In some places, factors of $c$, $G$, $\hbar$ are omitted.

\subsection*{Teleology and Terminology}

Different workers mean different things by ``black hole.''  Astrophysicists generally have in mind very compact objects which would explain certain phenomena; they are often not directly concerned with whether an event horizon in a strict sense exists, although this question is increasingly coming into the astrophysical literature (usually as the question of whether an object ``has a surface'').  At the other end of the spectrum, for mathematical relativists, the existence of the event horizon is central.  It will be clearest to begin by attempting to formalize the concept:

\begin{quote}
A {\em black hole} is a region of space--time from which causal signals can never escape; the {\em event horizon} is the boundary of the black-hole.  
\end{quote}
\noindent
These concepts are problematic, because of the words ``escape'' and ``never.''  Since there is no fixed sense of space in general relativity, the only invariant way of defining escape is a limiting one of ``escape to infinity.''  Even more severe are the consequences of 
``never.''  To verify that one has a black hole, according to this standard formulation, one would have to know the entire future development of the space--time, including over cosmic scales. This is the {\em problem of teleology}~\citep{Carter:1979,Booth:2005qc}.

One might hope to get around this by finding some local criterion which would {\em imply} that one had found an event horizon or a black hole.  However, such hopes turn out to be misdirected.  While one can indeed show that under certain assumptions event horizons will form, simple arguments also show that the locations of these cannot be fixed from local data, and indeed the location of the horizon could always potentially be altered by the {\em later} evolution of the space--time.  
The problem of teleology is central --- at least if we keep the usual definition of ``black hole.''

Because this issue is so severe, we should properly be open to reconsidering the definition.  But the present usage is firmly embedded, and until such time as a compelling alternative appears, it is pointless to try to change it.  Still, to carry the discussion forward, it is necessary to give some indication of what appears to be the problem and how we might resolve it, which involves contemplating some sort of change in what is to be called a black hole.

It seems clear that the root of the problem is an over-idealization, bound up with considering arbitrarily large times and distances.  Any practical reference to black holes must come down to considering times which are ``sufficiently  long'' but not actually infinite, and likewise ``sufficiently large'' regions of space.  These in turn show us that {\em a practical notion of black holes should be linked to systems which are complete enough to be considered  isolated}, in both time and space.  For the moment in will not be necessary to be more specific; we will accept that in some circumstances it is possible to characterize such systems and to use this sense of isolation to define (perhaps only up to a 
well-understood ambiguity) black holes and event horizons.

We therefore need terminology which allows us to specify just what system is under discussion, and what assumptions are going to go into modeling it mathematically.

I shall reserve the term {\em Universe} for the physical world.  By {\em a space--time} I mean the conventional mathematical concept of a manifold $\MM$ equipped with a Lorentzian metric $g_{ab}$.\footnote{Space--times are always assumed to be oriented, time-oriented (there is a globally sensible distinction between past- and future-directed causal vectors), and strongly causal (there are no curves which come ``arbitrarily close'' to violating causality --- see \citet{Penrose:1972} for formal definitions.}  We are usually interested in a space--time as {\em modeling} a portion of the Universe, most often a relatively small portion, such as an object we think is, in a suitable sense, a black hole.  When we do this, we should be as precise as possible about the purposes for which the model is supposed to be valid (which physical quantities we intend to be calculable and how accurately), and its extent in time and space.  Given the problem of teleology, it it especially important to consider how sensitive the predictions might be to any boundary conditions we assume, and how well justified those assumptions are.  

It will be important to distinguish between {\em complete} and {\em incomplete} models.\footnote{These terms have nothing to do with geodesic completeness.}  
By a complete model, I mean a space--time which is considered to model the entire physical system in question the entire region of interest; an incomplete model is a space--time which only models a portion (in time and space) of the system.
Suppose, for example, we have in mind a system with a central black hole which is surrounded by vacuum over some periods, but during others accretes matter.  It may be a good approximation for many purposes to model the exterior vacuum region by Schwarzschild solutions (and use some other solutions during the accreting periods).  In these cases the Schwarzschild solutions are incomplete models, but the entire patched-together space--time may be a complete model, if it covers the system over all times of interest.

In an example like the one just given, there may well be an event horizon associated with the complete model space--time.  The location of such a horizon cannot be determined from the incomplete models alone.  For instance, while a Schwarzschild solution of mass $M$ might form one of the incomplete models, the event horizon of the complete model will generally not lie at $r=2M$, even within the portion on which the incomplete model is valid.  It will be helpful to have a term for the surface $r=2M$ in this case, or more generally for the surface
where the event horizon would lie, were the black-hole space--time for the incomplete model valid; I will call this a {\em stand-in horizon}.

In much of the literature, the term ``event horizon'' is used for what I have called a 
stand-in horizon; also, typically Killing horizons are of interest as stand-in horizons.

\section{Black Holes and Causal Structure}

In this section, I will review the ideas leading up to the definition of black holes, the relation with singularities, and some of the fundamental theoretical results.

\subsection{Trapped Regions and Singularities:  Weak Cosmic Censorship}

While the history of the concept of black holes is complex and nuanced~\citep{Israel:1987ae}, a decisive step towards the modern view took place with the work of \citet{Oppenheimer:1939ue}.  These authors considered the indefinite gravitational collapse of a simplified (spherically symmetric, zero pressure) model of a star, and uncovered two key features:  the development of singularities and of a trapped region.
While the two features are known to be linked (via some of the singularity theorems),
we have at present only an incomplete 
understanding of the relationship between the two.  
It will help to begin by discussing the concepts individually.

We start by distinguishing between {\em local} and {\em global} notions of trapping.  The definitions of black holes and event horizons are global:  they are given in terms of whether causal signals can escape ``to infinity,'' which formalizes the idea of a region of space--time closed off from the inspection of distant observers. 

The notion of trapping which appears in the singularity theorems, however, is a local one, and it is an assumption, not a conclusion.  A {\em trapped surface}  is  a spacelike two-surface, both of whose null normals are converging.  In other words, a trapped surface is one from which a flash of light emitted ``outwards'' would (at least initially) decrease in area.
Notice that this concept does not require us to make any assumptions about the existence of an asymptotic regime where very distant observers are considered to lie.  However, if such an asymptotic regime does exist, then one can show under mild positivity-of-energy assumptions that the trapped surface must lie within an event horizon.

(That trapped surfaces lie within event horizons but can be studied locally has led to a great deal of work on them, and the related concepts of apparent, isolated and dynamical horizons~\citep{Hawking:1973uf,Ashtekar:2004cn}.
These concepts, while of interest, do not in general situations capture the sense of an event horizon, as a barrier to distant observers' examination.
Precisely because trapped surfaces lie behind event horizons, we hope and expect never to find them observationally!  And easy examples (e.g.~\citep{Booth:2005qc}; see also~\citep{Williams:2008} for more sophisticated work) show that in general trapped surfaces do not encode much of the geometry of the horizon, unless one makes teleological assumptions.)

Now let us turn to singularities.

What, precisely, do we mean by a space--time singularity?
Even in cases where the metric is known explicitly, it can be quite difficult to analyze the structure of a singularity, or even to verify that there is a singularity.  If one can show that freely falling observers would, in a finite interval of proper time, perceive divergent curvature tensors, for example, one has in a clear sense a curvature singularity, but it is not at all clear that these are the only possible sorts of singular behavior.  (See Clarke's book~\citep{Clarke:1993} for general results.)  And usually we only know the metric explicitly on an initial-data set, or in some region, and the explicit determination of the evolution is beyond our present technical capabilities.

For this reason, most of the strong results we have use an indirect characterization of singularities:  their existence is signalled by the presence of causal geodesics whose affine parameters do not take all real values --- incomplete causal geodesics.  One must then make sure that the incompleteness really does indicate a singularity, and not simply that some points have been left out of an otherwise non-singular space--time.  For this reason, the singularity theorems typically are phrased in the form, ``If certain conditions hold, then space--time cannot be causally complete.''

For instance, the 1970 singularity theorem of \citet{HP1970}  asserts that space--time cannot be causally complete if the following conditions hold:  (a) there is a trapped two-surface; (b) a positvity-of-energy condition holds
($R_{ab}t^at^b\leq 0$ for all timelike vectors $t^a$); (c) there are no closed timelike curves; and (d) a certain genericity condition holds (each causal geodesic contains at least one point along which the tangent is not an algebraically special vector for the Riemann curvature). 
The important points are that:  we do expect that such situations can occur; and, given any set of data satisfying  the hypotheses, all ``sufficiently close'' data will as well.  Thus the occurrence of singularities is a stable phenomenon (but the theorem does not tell us that the character of the singularities is stable).

So if the hypotheses of the theorem hold certain singularities will develop; if additionally there is a ``nice'' asymptotic region in space--time, those singularities will lie behind an event horizon; we say the horizon {\em clothes} them.  However, we do {\em not} know that those singularities are the only ones which may occur; in particular, we do not know that all singularities must lie behind an event horizon.  One which did not, that is, which would be visible to distant observers, would be called {\em naked}.  The question of whether naked singularities can exist is usually considered the most important open issue in classical general relativity:  it is the

\begin{quote}
\begin{center}
{\bf Weak Cosmic Censorship Question.}
\end{center}\nobreak
In classical general relativity, for physically realistic matter, in generic circumstances, must any singularities which develop lie behind an event horizon?
\end{quote}

This is somewhat imprecise, as the notion of what counts as ``physically reasonable'' has not been specified, nor has the notion of genericity, and also implicit in the formulation is the notion that the space--time has a regular enough asymptotic region that it makes sense to define an event horizon.  Nevertheless, it does capture the idea.  

There is no definitive progress, one way or the other, on this problem.  Penrose's article~\citep{Penrose:1999} is to be recommended as a prolegomenon.  I will make only a few comments:

First, it is quite surprising to conjecture a link between singularities, which are local, and event horizons, which are global.  For this reason, it has been thought that if weak cosmic censorship holds, it will do so as a special case of {\em strong} cosmic censorship, which should forbid the development of locally naked singularities.

Second, the genericity condition makes it very hard to approach this question via the study of exact solutions.  No single solution could provide a negative answer to the question; one would need to show that naked singularities persisted for an open set of data sufficiently close to the solution.  It should be emphasized that a really rigorous argument, and not (say) simply a linearized analysis would be required.
On the other hand, a generic numerical space--time with physically reasonable matter with a naked singularity could be strong circumstantial evidence for the failure of cosmic censorship.

Similarly, proof that a specific class of space--times (whose initial data lie only on a thin set in the space of all physically acceptable data) can  only develop clothed singularities does not settle the question in the affirmative.  However, such results could give clues about approaching the problem; the recent work of \citet{Ringstrom:2008} has generated excitement among those approaching the problem from the point of view of partial differential equations.

For the purposes of this paper, there are two points to keep in mind about naked singularities:
\begin{description}
\item[(The bad news)]  Because we have so little definitive progress on the question of weak cosmic censorship, we know little about what a naked singularity (were one to form) would look like.

\item[(The good news)]  A naked singularity (were one to form) would, unlike an event horizon, be by definition discernible by distant observers at finite times.  Thus naked singularities, whatever they might be, should be distinguishable from black holes.

\end{description}

\subsection{Hypotheses on the Asymptotic Regime}

In order to formalize the notion of a black hole, we must say what it means for causal signals to escape from a system. 
 To do this we must define a regime which counts as ``very far away'' and is accessible by causal curves.  There is in fact no single accepted notion.  Mathematically, this can be thought of as an issue of the regularity one wants to  hypothesize in the asymptotic regime; physically, one is deciding what exactly should qualify as a ``system.''  In general relativity, the questions of what the system is and what its asymptotics are must be addressed together.

For most purposes, one does not want to have to consider the whole Universe in order to analyze a system potentially containing a black hole; at the same time, one would hope that the precise boundaries of the system are not too important and one can substitute some sort of idealized boundary conditions for (what would be more accurate) a precise specification the relevant data and its uncertainties at the boundary.  One expects this to be largely true, but there will be places where it will be important to recall that we are really working with idealizations which cannot be pushed too far.  Again, we are here distinguishing the physical {\em Universe} from the mathematical {\em space--time}, which signifies a model of a portion of the Universe subject to some idealizations.

With few exceptions, one is interested in modeling black holes within systems which can be viewed as isolated.  The relevant model of the asymptotic regime is then that first given by Bondi, Sachs and coworkers~\citep{Bondi:1962,Sachs:1962}.
While it was 
originally developed by finding an appropriate class of asymptotic coordinate systems with which general radiating systems could be studied, \citet{Penrose:1964} showed that it could be recast neatly and powerfully in terms of the space--time admitting a ``conformal boundary.'' 

While the end result of the Bondi--Sachs treatment is easy to work with, the logic justifying it includes some fine points.  I am going to quickly review it, with two aims:  first, I want to emphasize that it is not at all clear how closely parallel a structure would exist  in other dimensions, so one should be cautious of assuming that the theory of black holes (or isolated systems generally) in higher dimensions will follow the same pattern as in four;\footnote{For an important feature which does not generalize, see~\citet{Hollands:2004ac}.} second, I want to raise a mild concern about one assumption common in some of the literature.  I will not give the details of the calculations or the differentiability assumptions, for which see \citet{Penrose:1986ca}.

I will say a space--time $(\MM ,g_{ab})$ {\em admits a future null conformal infinity} $\scrif$ if: (a) $\MM$ embeds as the interior of a manifold $\overline\MM$ with boundary $\scrif$; (b) there exists a smooth function $\Omega$ on $\overline\MM$ with $\Omega >0$ on $\MM$ and $\Omega =0$ on $\scrif$
and ${\hat N}_a=-{\hat\nabla}_a\Omega$ is non-zero on $\scrif$; (c) the conformally rescaled metric ${\hat g}_{ab}=\Omega ^2g_{ab}$ extends smoothly to a non-degenerate metric on $\overline\MM$; (d) each point of $\scrif$ contains the future, but not the past, end-points of null geodesics in $\MM$.\footnote{Here $ {\hat\nabla}_a $ is the covariant derivative operator with respect to ${\hat g}_{ab}$; subsequently, we will use the convention that hatted quantities have their indices raised and lowered with ${\hat g}^{ab}=\Omega ^{-2}g^{ab}$ and ${\hat g}_{ab}$.}

That ${\hat g}_{ab}$ should be regular at $\scrif$ is in fact a powerful constraint.  For instance, the Ricci scalars of the 
original and rescaled metrics are related by
\begin{equation}\label{Rsc}
R=\Omega ^2{\hat R} -6\Omega{\hat\nabla}_a{\hat\nabla}^a\Omega +12{\hat N}_a{\hat N}^a\, .
\end{equation}
Einstein's equation gives us $-R+4\lambda =-8\pi G T_a{}^a$, where $\lambda$ is the cosmological constant of the space--time (not the Universe).  This
cosmological constant would be relevant {\em only} if we wanted to treat the model as applicable on cosmological scales, and we usually do not: in the Bondi--Sachs scheme we assume $\lambda =0$.  And
because points on $\scrif$ are reached by going out to infinity along null geodesics, we generally expect that the only matter fields which can survive in this limit are null radiation, which has $T_a{}^a=0$.  In this case,
we have 
${\hat N}_a{\hat N}^a=0$ at $\scrif$, and
$\scrif$ is 
a null hypersurface.
As such, it is foliated by null geodesics (of the rescaled metric), called its {\em generators}.

Each point on $\scrif$ can  be thought of as the set of null geodesics in the physical space--time which terminate at the point.  Since there is a five-parameter family of null geodesics (neglecting the two additional choices needed to fix affine parameterizations of them), and $\scrif$ is three-dimensional, we expect each point on $\scrif$ to correspond to a two-parameter family of null geodesics.  
Each such family can be identified with an asymptotically plane-fronted wave-front.
Note that each null generator of $\scrif$ corresponds to a one-dimensional family of such wave-fronts, at a succession of later and later times.  We may interpret this as saying that all of these wave-fronts have the same asymptotic direction.
We then make the natural assumption that $\scrif$ is topologically
$S^2\times\R$, the angular variables corresponding to the different possible directions for the asymptotic wave-fronts, and the $\R$ factors to the null generators, that is, the different possible retarded times.
 (If one is willing to assume asymptotic simplicity --- see below --- then one can prove that $\scrif$ must have this topology~\citep{Penrose:1965,Geroch:1971}.)  

It is often convenient to think of $\scrif$ as a bundle over $S^2$, projecting along the generators; in this context, a section of the bundle is called a {\em cut} of $\scrif$.  The base space $S^2$ is then naturally the set of asymptotic directions in which null geodesics might escape.  However, it is not at all obvious that this base space has structure beyond that of a smooth point-set.    But in fact it turns out that this $S^2$ has well-defined conformal structure, and this is critical to some of the deepest results.

We know already that any cut of $\scrif$ has a conformal structure, simply by restricting the metric ${\hat g}_{ab}$ to the cut.  What is not obvious is that flowing along the generators of $\scrif$ should preserve this conformal structure.  The condition that this should be the case is that the null tangent ${\hat N}^a$ should be {\em shear-free}, that is ${\hat\nabla}_{(a}{\hat N}_{b)}$ should be pure trace.  However, by a calculation like that leading to \eqref{Rsc}, this turns out to follow from the (very weak) assumption  that $\Omega (T_{ab}-(1/4)g_{ab}T_c{}^c)$ vanishes at $\scrif$, and we henceforth assume this holds.

There is further important structure, which again is not obvious.  In general, if one only knows the conformal class of a metric, its null geodesics are well-defined only up to reparameterization:  in particular, they do not have a well-defined affine structure.  This would appear, at first blush, to be the case for $\scrif$.  However, it turns out that the null generators of $\scrif$ {\em do} have well-defined affine structures.  This arises from the invariance of the quantity ${\hat g}_{ab}{\hat N}^c{\hat N}^d$, sometimes called the {\em strong conformal geometry} of $\scrif$.  Picking any one metric structure for the base $S^2$, we may pull this back to $\scrif$, and this fixes a scale for the tangent ${\hat N}^a$ to the generators.  While the scale may change with the metric chosen, it can only do so by a factor which is constant along each generator. 

We may get still more, for we may restrict the metric on the base $S^2$ to be that of a unit sphere, compatible with its conformal structure.  There is only a three-dimensional family of such choices, which is naturally identifiable with the set of unit timelike vectors.  Therefore each such choice amounts to an asymptotic choice of timelike unit vector.

What this means is that, for each asymptotic direction, there is a well-defined sense of retarded time $u$ along the corresponding generator, transforming naturally with the choice of asymptotic unit timelike vector, and otherwise free only up to the addition of a (generator-dependent) term.
(This freedom is the famous {\em supertranslation ambiguity}.)  In particular, it is meaningful to say whether the generator extends infinitely far, in either the future or the past.  To model 
black-hole space--times, we shall assume that the generators are infinitely long in both directions.  (One might consider relaxing the requirement that they be infinitely long to the past.)

Much more can be deduced from this, including {\em Sachs peeling,} which describes the asymptotic fall-off of the curvature, and this formalism is also the basis for the Bondi--Sachs energy--momentum~\citep{Penrose:1986ca} and twistorial angular momentum~\citep{Helfer:2007}.

{\em A Comment on Weak Future Asymptotic Simplicity.   }
The treatment that I have given here is almost completely standard.  However, there is one assumption which is commonly made.

A space--time $(\MM ,g_{ab})$ is {\em future asymptotically simple} if it admits a future conformal null infinity and if additionally
{\em every} null geodesic in $(M,g_{ab})$ acquires a future end-point on $\scrif$.

Now, it is well-known that it is overly restrictive to impose future asymptotic simplicity; for example, in Schwarzschild there are null geodesics which orbit at $r=3M$ and do not escape.  
For this reason, it has become common to consider {\em weakly future asymptotically simple} space--times, which are those which admit an open set isometric to an open neighborhood of $\scrif$ in an auxiliary future-simple space--time.  (More precisely, isometric to an open {\em deleted} neighborhood of $\scrif$, that is, deleting $\scrif$ itself.)

It seems to me that perhaps the term ``simplicity'' occurring in these definitions has given the impression that they correspond to expected situations.  While it is certainly of interest to try to investigate the future limits of null geodesics, it is not {\em a priori} clear that the set of those future limits which correspond to the idea of escape from a given system should have the property of weak asymptotic future simplicity.  So I would suggest that until this point is clarified, it would be prudent not to assume that all isolated systems of interest are weakly future asymptotically simple. 

\subsection{Black Holes, Event Horizons, and Causal Structure}

In a space--time $(\MM ,g_{ab})$, we may define the {\em causal past} of any subset $S\subset \MM$ as
\begin{equation} 
  J^-(S)=\{  p\in \MM\mid\mbox{there is a causal curve from }p\mbox{ to some point in }S\}\, .
\end{equation}
(A causal curve, if smooth, is one whose tangent is everywhere a future-directed timelike or null vector.  For the non-smooth case see~\citep{Penrose:1972}.)  Note that this depends only on the conformal class of the metric.  Thus if $(\MM ,g_{ab})$ admits a future null conformal infinity $\scrif$ as defined above, we may define
\begin{equation} 
  J^-(\scrif )=\{  p\in \MM\mid\mbox{there is a causal curve from }p\mbox{ to some point in }\scrif\}\, ;
\end{equation}
this will be the set of events in $\MM$ from which at least one causal curve escapes to the asymptotic regime $\scrif$.  If $\scrif$ has been defined so that it includes all possible means of ``causal escape'' --- usually, this is taken to mean that the space--time is weakly future asymptotically simple, but see the comment at the end of the last subsection --- then we may interpret
\begin{equation}
 \B =\MM -J^-(\scrif )\, ,
\end{equation}
if it exists, as the {\em black-hole} region of the space--time.  The {\em (future) event horizon} is the boundary of this:
\begin{equation}
\Hf =\partial\B\, .
\end{equation}
(Most authors assume, as part of the definition of a black hole, that the space--time is {\em globally hyperbolic} (see, e.g.~\citep{Penrose:1972}), which essentially means that it admits a good initial-data surface.  I have omitted this because it is bound up with the question of cosmic censorship~\citep{Penrose:1980}.)

These points should be emphasized:  
\begin{itemize}
\item
We are only entitled to interpret $\B$ as a black hole if we are satisfied that $\scrif$ represents all possible means of causal escape.
\item
The definition of the black hole (and that of the event horizon) is {\em highly time-asymmetric}.  The time-reverse concept would be a {\em white hole} (given by $\MM -J^+(\scrip )$ in an obvious notation), a region of space--time which incoming causal signals could not penetrate --- something quite different.
\item
The black hole and its event horizon are determined {\em highly nonlocally}, in terms of whether escape is ever possible.  This has been one of the most difficult features to work with in the theory.
\end{itemize}

There is sometimes a tendency to think that because black holes are expected in many cases to settle down to stationary states  this time asymmetry is not very important.  However, it is crucial for some purposes in surprising ways --- for instance, it was Hawking's appreciation of this which led him to predict black-hole radiation, when others had intuitively ruled it out based on stationarity arguments.  It is worthwhile, when dealing with black holes, to make one's arguments about which parts of the space--time are expected to be (approximately) stationary carefully.

\subsubsection*{Structure of the Event Horizon}

The causal differential topology of the event horizon $\Hf$ is elegant.  For proofs of the following, see~\citet{Penrose:1972}:

(a) It is not hard to see that $\Hf$ must be {\em achronal}, that is, no timelike curve can join one of its points to another.  
(If $p,q\in\Hf$, if there were a timelike curve from $p$ to $q$, then, displacing $q$ slightly to a point $q'\in \MM -\B$ to the past of $q$, we could get a causal curve from $p$ to $q'$ and then escaping to $\scrif$, a contradiction.)
(b)
The event horizon must also be a topological three-manifold which is locally the graph of a Lipschitz function (that is, has bounded difference quotients).
(c)
Any point $p\in\Hf$ lies on a null geodesic segment (not necessarily unique) on $\Hf$ extending to the future of $p$, and this segment lies on $\Hf$ as far as the segment extends.
(d)
No two null geodesic segments on $\Hf$ can end at the same point, unless they are the same segment in a neighborhood of the point.

These results again point up the highly time-asymmetric character of the horizon:  it may acquire new generators as one moves to the future but, once a null geodesic joins the horizon it can never leave.  New generators join at {\em caustics,} and it is of great interest to understand just how this happens and what observational properties the neighborhood a black hole where a caustic forms might have.

I have pointed out that the (standard) arguments so far give what is usually regarded as a rather low degree of regularity for event horizons:  they are locally Lipschitz.  Is this really as much as can be said?

As far as the purely mathematical question goes, not much more regularity can be expected in general.  (See~\citet{Chrusciel:2001,Chrusciel:2002} for slightly better results.)  The precise statements of the results are technical (for example, horizons are differentiable almost everywhere in the sense of measure theory but may well fail to be differentiable over all open sets~\citep{Chrusciel:1998}).   The cleanest general result is one of \citet{Beem:1998}, who show that differentiability fails precisely at places where new generators join.

But are such mathematical oddities really relevant physically?  In order to investigate this, let us suppose we are modeling a system with Bondi--Sachs asymptotics. 
The non-black region is the interior of $J^-(\scrif )$, and we may think of forming this by a limiting process.  Let $Z$ be any cut of $\scrif$, and let $\scrif _Z$ be the portion of $\scrif$ at or prior to $Z$.  Then $J^-(\scrif )=\bigcup _{Z\uparrow +\infty} J^-(\scrif _Z)$, where the union is over later and later cuts.   We may take this union over a family of smooth cuts.

Each $\partial J^-(\scrif _Z)$ will be, in the vicinity of $Z$, a smooth null hypersurface meeting $Z$ orthogonally.  As we follow it into the past, it may develop singularities, associated with what is called the {\em cut locus} (this use of ``cut'' is different from $Z$; unfortunately, both uses are firmly embedded) --- the family of points at which its generators begin to cross; at these cut points, some of the null geodesics, followed into the past, would move to the chronological past of $Z$ and hence leave $\partial J^-(\scrif _Z)$.  However, such cut points would arise by solving for the intersection of two null generators of $\partial J^-(\scrif Z)$, and generically (that is, at points where neither generator is conjugate to $Z$) this variety will be a smooth two-surface.  

Also one would expect that for generically positioned smooth $Z$ only finitely many such cut loci will occur as one follows $\partial J^-(\scrif _Z)$ inward to a neighborhood of any point on $\Hf$.  On the other hand, as we move $Z$ later and later, it is quite possible that the number of such cut loci increases, and that they accumulate, these $\partial J^-(\scrif _Z)$ accumulating in an irregular way as the horizon is approached.

It thus seems to me likely that the question of how irregular practical approximations to the horizon are in realistic models is a reflection of how increasingly complex the system of cut loci associated to the cuts $Z$ become as $Z$ moves towards the future.

\subsection{The Area Theorem}

The famous Area Theorem is a key conceptual result.
It asserts that, if the weak energy condition holds ($T_{ab}l^al^b\geq 0$ for all null vectors $l^a$), then the areas of cross-sections of the event horizon increase as the sections move towards the future.

A ``physicist's'' statement and proof of this theorem was first given by Hawking in 1972~\citep{Hawking:1971vc}, and I shall outline this below.  While this does provide the template for a rigorous treatment, the technicalities in achieving that are formidable.  Given the low regularity of the event horizon, it is not even clear that there are many cross-sections whose areas can be defined; also the physicist's argument is overly cavalier in its use of differential inequalities for surface area elements and in its treatment of caustics.  
A proof addressing the necessary regularity issues --- and indeed clarifying some interpretational points --- was only given in 2001, by Chrus\'ciel, Delay, Galloway and Howard~\citep{Chrusciel:2001}.

Here is the ``physicist's argument'':  Let $l^a$ be the tangents to the affinely parameterized null generators of $\Hf$.  (For now, we only consider points at which there is a unique tangent.)  Then the {\em Raychaudhuri equation} (derived from the geodesic deviation equation) is
\begin{equation}
l^a\nabla _a\rho =\rho ^2+|\sigma |^2 +4\pi G T_{ab}l^al^b\, ,
\end{equation}
where $\rho$ (real) and $\sigma$ (complex) are the convergence and shear of $l^a$.  
Assuming the {\em weak energy condition} $T_{ab}l^al^b\geq 0$,
this shows that $l^a\nabla _a\rho\geq \rho ^2$.  
Any function satisfying this inequality will diverge to infinity within a finite parameter range, if 
$\rho$ ever becomes positive. 
But if this were to happen either the generator would be incomplete (which would correspond to a singularity) or it would develop a conjugate point, which is not allowed because that would require a null generator to enter the black-hole region.  Thus, excepting the possibility of an incomplete generator, we must have $\rho\leq 0$ everywhere.

Since $\rho$ is the convergence of $l^a$, it can be interpreted as $-(1/2)(\Lie _l dA)/dA$, where $dA$ is the area element transverse to the generator.  Thus $\rho \leq 0$ is a statement that the area element cannot decrease as we flow forward along generators.  Since generators can never, as we move forward in time, leave $\Hf$, but new generators can join, we expect the appearance of caustics only to lead to further increases in the area. This completes the ``physicist's argument.''

\begin{wrapfigure}{r}{0.4\textwidth}
\vspace{-2em}
  \begin{flushright}
{\includegraphics[width=.35\textwidth]{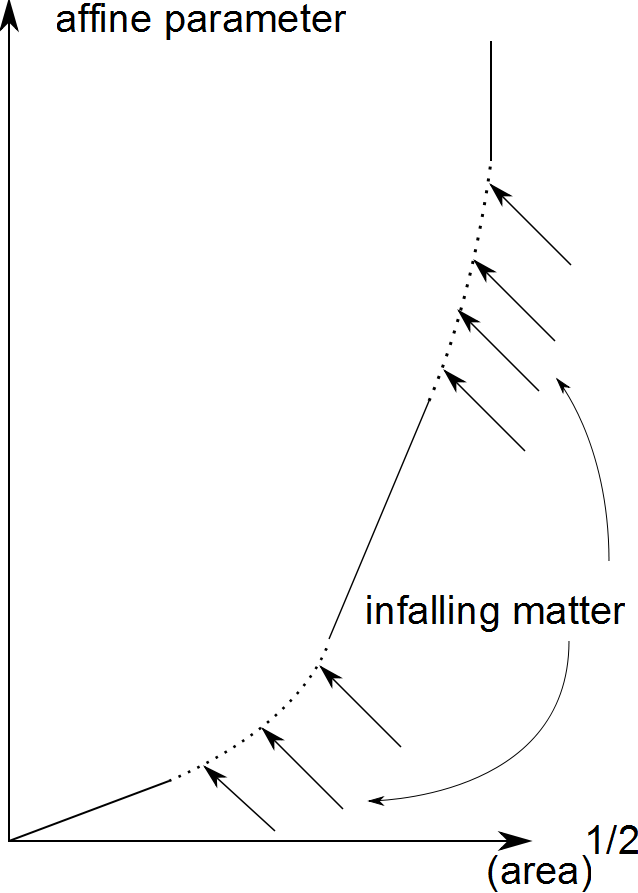}}
\makebox[.35\textwidth]{$\ $ }
\vspace{-2em}
\parbox{.35\textwidth}{1. The teleological character of black-hole area.  The square root of an area element grows when no matter is falling in, in {\em anticipation} of a later infall.  In fact, the growth is linear in vacuum (if there is no shear) and sublinear during infall.}
\end{flushright}
\vspace{.5em}
\label{fig:telarea}
\end{wrapfigure}

Note the teleological character of the argument:  the convergence can never, {\em in the future}, become positive, and therefore the Raychaudhuri equation and the weak energy condition imply that it must be non-positive {\em at earlier times.}
In fact, if we push this argument a little further, we see that what we have is a differential equation with a {\em final}, rather than an initial, condition.  We have seen that $\rho$ is a monotonically increasing function, bounded above, on each generator, and so its limit must exist.  But this limit cannot be negative, for that is not allowed by the Raychaudhuri equation.  Thus $\rho$ must tend to zero, and we have the integral equation
\begin{equation}\label{rhoint}
\rho (s)=-\int _{s}^\infty ( (\rho ^2+|\sigma |^2 +4\pi G T_{ab}l^al^b)(\as ) )d\as\, ,
\end{equation}
where the bizarre range of integration, from the affine parameter value $s$ in question {\em forward} in time, makes manifest the teleological character of the situtation.

Imagine, for example, a spherically symmetric black hole forming from an initial infall of matter to a vacuum region; say this matter has crossed the event horizon for affine parameters $s_1\leq s\leq s_2$.  In general, the event horizon will have formed {\em before} the matter got there, in anticipation of its arrival.  Similarly, if during a later interval $s_3\leq s\leq s_4$ more matter falls in, then during the no-infall period $s_2\leq s\leq s_3$ the event horizon will expand in anticipation of that infall.  In fact, the growth of the square root of the area element can be linear in the no-infall periods but only sublinear in the infall periods; see Fig.~1.\footnote{This sort of behavior was noted by~\citet{Carter:1979} and quite explicitly by~\citet{Booth:2005qc}.}  It is this behavior which the Area Theorem reflects.

\subsection{Stationary Black Holes}

The Area Theorem is the strongest general, dynamical, result on black holes.  
The problem of developing further dynamical results in full general relativity is largely open, although there is a large body of perturbative results and also an increasing body of numerical work.
There does exist, however, important work on stationary black holes.

This material may be roughly divided into a family of results and expectations which is usually regarded as adequate for most physical purposes, and the serious mathematical problems involved in justifying them.  
I am here going to sketch the results and the (sometimes rough) arguments given for them.  There is much rigorous work in this area, and it is very active, but giving an account on it
would involve many qualifications and technicalities.
Among older works,
Carter's review article~\citep{Carter:1979} is a good introduction to the material, and of course many important ideas are in \citet{Hawking:1973uf}.  
A recent sampler, with references, will be found in section 3 of the useful paper by Chrus\'ciel, Galloway and Pollack~\cite{Chrusciel:2010fn}.

There is a general expectation that in many cases a black hole will settle down to be well approximated, at least at large distances, by a stationary space--time, that is, one which possesses a Killing vector $\xi ^a$ which is timelike at large distances.   
(That $\xi ^a$ need not be timelike everywhere will be discussed in more detail shortly.)
We assume that at large distances $\xi ^a$ approaches a unit future-directed vector.

One would expect that in order to be stationary, the space--time must have a significant degree of symmetry, for typically inhomogeneities lead to multipole moments which result in gravitational radiation, which is not compatible with stationarity.  One possibility is that the space--time is not only stationary but in fact static, that is, possesses a time-reversal symmetry.  Such a situation is usually regarded as very specialized; I shall not discuss it further.  Another possibility is that the space--time is axisymmetric, that is, possesses a second Killing field $\phi ^a$, with closed spacelike orbits, commuting with $\xi ^a$.  (If $\phi ^a$ did not commute with $\xi ^a$, their commutator would be another Killing field and one would be in a more specialized situation.  Under mild assumptions one can prove that the two must commute \citep{Carter:1970ea,Carter:1979}.)  One can in fact prove, subject to certain assumptions, that a non-static stationary black-hole space--time must be axisymmetric in this sense~\citep{Hawking:1973uf,Chrusciel:2010fn}.

When suitable assumptions are made on the matter fields, one may hope to classify the stationary axisymmetric space--times.  The ``no-hair'' principle asserts that the only Einstein--Maxwell such solutions are those of the Kerr--Newman family; in particular, a stationary, axisymmetric vacuum solution is expected to be a Kerr solution.

{\em Ergoregions and Energy Extraction.}
The portion of space--time exterior to the black hole in which $\xi ^a$ is not timelike is the {\em ergoregion;} the boundary where $\xi ^a$ makes the transition to timelike is the {\em stationary limit surface}.
It can be shown that at any event in the exterior the two-plane spanned by $\xi ^a$ and $\phi ^a$ is timelike (becoming null precisely at the event horizon), and so even a point in the ergoregion will have a timelike Killing field in its vicinity.
However, the discrepancy between this field and $\xi ^a$
gives rise to the possibility of extracting significant amounts of energy from a highly rotating black hole.

If  a particle's four-momentum is $p_a$, its $\xi ^a${\em -energy} is $E=\xi ^ap_a$.  It is not hard to show that this is conserved for a freely falling particle, and coincides with the usual energy measured in a frame defined by $\xi ^a$ in the asymptotic region.  But now suppose the particle falls into the ergoregion, and splits into two, locally conserving energy--momentum.  Because the vector $\xi ^a$ is space-like there, it is possible for the daughter particles to have $\xi ^a$-energies $E_1>E$, $E_2<0$.  Then the first of these may escape to infinity with an augmented energy (while the second, having $E_2<0$, cannot escape  to the asymptotic regime).  In this way {\em energy may be extracted from a black hole}; this is called the {\em Penrose process}~\citep{Penrose:1965}.  

The real point of the argument is not, however, the specific process, but the possibility of energy extraction.  Suppose, for example, one very gradually extracted energy from a Kerr solution, giving it time to equilibrate.  Then, by the no-hair principle, it would pass through a family of Kerr solutions, and 
for slow enough extractions one would expect this could be done adiabatically, that is, without ever deviating substantially from the Kerr family.  Now, only a more detailed analysis could define and track the event horizon; each Kerr solution has rather what I called at the end of the introductory section a stand-in horizon. 
Still, for the adiabatic process just described, it seems plausible that the event horizon should track the stand-in horizons.  Then the Area Theorem would provide a limit on the energy extraction, but it is hard to see that that there should be any other fundamental limit.  For a Kerr black hole with mass $m$ and specific angular momentum $a$, this limit is
\begin{equation}
  m(1-2^{-1/2}(1+(1-a^2/m^2)^{1/2})^{1/2})\, .
\end{equation}
If $a$ is a significant fraction of $m$, this is a significant fraction of $m$ as well, approaching $m(1-2^{-1/2})\simeq .29 m$ as $a\uparrow m$.  Thus if astrophysical black holes acquire large specific angular momenta, colossal energies could be extracted from them.

{\em Angular Velocity and Surface Gravity.}
The Killing fields $\xi ^a$, $\phi ^a$ must be tangent to the event horizon $\Hf$ (for it is a geometric invariant).  The {\em rigidity theorem}~\citep{Hawking:1971vc,Hawking:1973uf} asserts that there is a Killing field which is tangent to the generators of $\Hf$, and
(assuming there are no symmetries beyond those generated by $\xi ^a$ and $\phi ^a$), this field can only be a linear combination with constant coefficients of $\xi ^a$ and $\phi$.  We normalize it to be $\chi ^a=\xi ^a+\omega\phi ^a$.
Then $\omega \phi ^a$ can be interpreted as the {\em angular velocity} of the event horizon relative to infinity.  

On the horizon itself, since the generators are geodesics, we must have $\chi ^b\nabla _b\chi ^a=\kappa \chi ^a$ for some scalar $\kappa$, necessarily constant up the generators.  It is a remarkable fact that $\kappa$ {\em must in fact be constant over the horizon}, which can be shown by differentiating Killing's equation and using the {\em dominant energy condition} (the energy--momentum density $T_{ab}t^b$ measured by any observer with future-directed timelike tangent $t^a$ should itself be future-directed).
The constant $\kappa$ is called the {\em surface gravity} of the black hole.

It is important to note that the definition of surface gravity is non-local, for it requires the vector $\xi ^a$ to be normalized at infinity.  In fact, suppose a test particle rigidly co-rotates with a black hole, just outside the event horizon.  Then the surface gravity turns out to be the particle's acceleration divided by $u^a\xi _a$ (where $u^a$ is its unit future-directed timelike tangent)~\citep{Bardeen:1973gs}, so one might more properly call $\kappa$ the surface gravity, measured relative to infinity (for that is where $\xi ^a$ becomes a unit vector).

\citet{Wald:1984} gives a nice interpretation of surface gravity, for spherically symmetric holes.  Consider lowering a small mass slowly on a very light but inelastic string towards the event horizon from infinity.  As one lowers it, one can recover potential energy.  By working out the change in energy as the string is lowered, one can find the tension that must be applied by the holder of the string, and thus the applied acceleration to keep the string stationary.  One finds that as the end of the string approaches the horizon, the acceleration approaches the surface gravity.  (This argument works for strings lowered along the axis of spinning black holes, too.)

Another interpretation (valid for spherically symmetric holes, again) is as the logarithmic rate of change of the red-shift factor along radial null geodesics from the distant past to the distant future; this will be discussed in the next section.

The existence of a constant surface gravity is quite remarkable, and not fully understood.  It plays an essential role in ``classical black-hole thermodynamics,'' as well as in Hawking's quantum analysis of black holes.

{\em Energy and Angular Momentum.}
In general we do not have good definitions of energy--momentum or angular momentum in general relativity.  However, there are certain classes of situations for which we do have satisfactory results.  In particular, for space--times admitting Bondi--Sachs asymptotics these quantities can be defined at arbitatrary cuts of $\scrif$, and, if the space--time is stationary, there is a well-define sense in which they are constant (independent of the cut).  In the stationary axisymmetric case, the energy and angular momentum turn out to be given by the {\em Komar integrals}
\begin{equation}
  E=(8\pi G)^{-1}\oint \epsilon _{abcd}\nabla ^{[c}\xi ^{d]}
  \, ,\quad
  J=-(16\pi G)^{-1}\oint \epsilon _{abcd}\nabla ^{[c}\phi ^{d]}\, ,
\end{equation}
where the integrals are taken over a shear-free cut of $\scrif$.

In the case of vacuum exteriors, the surfaces of integration can be continuously deformed without affecting the results, and so it has been common to define the energy and angular momentum ``locally,'' or ``of the hole itself'' by deforming the surface to the event horizon.  However, while there is a good justification of this in the vacuum-exterior case, I would caution 
that in general the physical  interpretation of the Komar integrals at the horizon is obscure.
Some hint of this is seen in the famous factor of $2$ between the two, which signifies that the 
Komar integrals do {\em not} give the
conserved quantities as linear functions of the associated Killing vectors, even at $\scrif$.  
(That is, the conserved quantity associated with $a\xi ^a +b\phi ^a$ is $aE-bJ$,\footnote{The minus sign arises from the convention used in converting the two-form for angular momentum to a vector.} but this is not the Komar integral of the Killing field $a\xi ^a+b\phi ^a$.)  The logic justifying the physical interpretation of the Komar integrals as energy and angular momentum applies {\em only} at infinity.

\subsection{Classical Black-Hole Thermodynamics}

The Area Theorem is clearly suggestive of the second law of thermodynamics, and soon after it was discovered \citet{Bardeen:1973gs} pointed out that there were other parallels.  
These authors in fact cautioned against taking the correspondence more literally, although very shortly thereafter several lines of thought came together which suggested that it was much more than formal.  Some of those later arguments, which depend on quantum effects, will be described in more detail in Section~3; here the treatment is classical.  It will be clearest to take the laws out of order.

The {\em second law} of classical black-hole thermodynamics is, as noted above, the Area Theorem.  There are, however, difficulties, some more serious than others, in making the correspondence with ordinary thermodynamics precise.  

First, the area of a black hole is a dimensionful quantity, whereas thermodynamic entropies are dimensionless (in units where Boltzmann's constant is one); Hawking's quantum analysis suggests that the entropy is $A/(4l_{\rm Pl}^2)$, where $l_{\rm Pl}$ is the Planck length.
Second, thermodynamic entropies are only defined when some sort of averaging process (perhaps an implicit 
coarse-graining) is used; by contrast, no such procedure is evident in the definition of a back hole's area.  
But the most important difference is that the area of the black hole is defined teleologically and is neither directly measurable in principle nor calculable from theory without teleological assumptions --- something which is very different from ordinary physical reasoning.

The {\em zeroth law} of thermodynamics (existence of temperature) corresponds, in black-hole theory, to the existence of surface gravity, that is, to the fact that the surface gravity of a stationary hole is not only defined but is constant on the horizon.
Classically, however, it is hard to see how this can be interpreted as a thermodynamic temperature (and its units are not those of temperature).  Hawking's remarkable quantum analysis, however, suggests that the incipient hole really does radiate at a temperature $T_{\rm H}=\hbar \kappa /c$.

Note however that the temperature and entropy
have in principle very
different bases in that the temperature is defined in terms of the {\em stand-in horizon}, whereas the justification for interpreting the area as an entropy requires using the {\em event horizon}.
Also, while Hawking's quantum analysis does suggest that a black hole really has a temperature in a conventional sense, it does not provide an {\em independent} explanation of just what the link between area and entropy is; the formula $A/(4l_{\rm Pl}^2)$ for the entropy is inferred, given the temperature, only by analogy with ordinary thermodynamics.  The problem of justifying the interpretation of black-hole area as an entropy is a major open one.  (This is one of the areas of work in quantum gravity theories.)

For the {\em first law}, we want a statement of conservation of energy in terms of physically satisfactory heat and work terms. 
We do not really have that.  One underlying difficulty is that, except in the simplest cases, we must have theory which allows relativistic changes of reference frames (for example, one could imagine a process which resulted in the direction of the black hole's energy--momentum changing), and
at present there is as yet no agreed fully 
{\em special}-relativistic thermodynamics (see e.g. \citep{DHH:2009}).  There is a related difficulty with angular momentum, in that one must be able to accommodate supertranslations.

There are two distinct versions of the first law discussed in the literature.  The first is the {\em equilibrium-state} version.  It only compares invariant information about two neighboring stationary black-hole states {\em individually}; it does not attempt to keep track of the additional invariant information needed to specify one of these {\em relative} to the other (asymptotically, the boost, rotation and supertranslation).  It is therefore really a differential relation on the invariant parameters (mass and total angular momentum) of the family of stationary black-hole space--times.

The equilibrium-state version is most straightforward in the stationary axisymmetric vacuum case (which is believed to comprise only the Kerr solutions), where it takes the form
\begin{equation}
\delta M=(8\pi )^{-1}\kappa\delta A +\omega\delta J\, ,
\end{equation}
which one may compare to
\begin{equation}
  \delta E=T\delta S -\delta W
\end{equation}
for an ordinary thermodynamic system, with $\delta W$ being the work done on the environment.  The analogy here is certainly very attractive.

An extension of this to the general axisymmetric stationary case has been given by \citet{Carter:1979}; it has the form
\begin{equation}  
  \delta M=\omega\delta J_{\rm H}+(8\pi )^{-1}\kappa \delta A
  +\delta\int T_{ab}\xi ^a d\Sigma ^b -(1/2)\int T_{bc} h^{bc} \xi _a d\Sigma ^a\, ,
\end{equation}
where $\Sigma$ is a spacelike three-surface joining the horizon to infinity, and $h_{ab}=\delta g_{ab}$ is the associated metric perturbation.  Two points are important here:  First, the diffeomorphism freedom has been used to adjust $h_{ab}$ so that the Killing vectors are preserved by the perturbation, and hence the relative boost, rotation and supertranslation of the two states are not encoded.  Second, the change in angular momentum $\delta J_{\rm H}$ which appears here is that defined by the Komar integral applied at the horizon.  Since we do not have a good justification for this except in the vacuum case, one is left with interpretational questions.

The second version of the first law is the {\em physical-process} form, and this aims to treat the true dynamical problem of passage of a system from one equilibrium state to another~\citep{Wald:1995yp,Gao:2001ut}.  However, so far, work on this has only treated the case where one need not consider perturbations of the background geometry, and in particular this means that one is restricted to situations where one does not have to consider changes in the Killing vectors (for example, to accommodate relative boosts, rotations or supertranslations).

Finally, the literature also contains discussion of a possible {\em third law}, which would hold
that one cannot by a finite sequence of even idealized processes reduce the temperature (i.e., surface gravity) of a black hole to zero.

\section{Holonomy and Black Holes}

Our present definition of black holes is teleological, which is counter to virtually all conventional physical considerations.  It is worthwhile to try to distance ourselves from this aspect of the theory, and try to identify what seem to be the prominent features of black holes for which we need not invoke teleology.  

In this section, I will sketch an approach to this problem which seems promising.  I will show that useful and potentially observable information is encoded in the infinitesimal holonomies relating the neighborhood of the event horizon to the regime occupied by distant observers.  More precisely, we imagine parallel propagation along an outgoing null geodesic from a point near the event horizon to a very distant point; the change in this propagation as we consider later and later null geodesics is (essentially) the infinitesimal holonomy.  

We will see that this quantity encodes certain {\em universal} features of the incipient hole, in that it is directly recoverable from signals emitted by matter falling towards the horizon, independent of the peculiar velocity of the matter or the point at which it will cross the horizon.  The verification of this universal behavior, and even more so, of its persistence for long times, provides strong circumstantial evidence that an event horizon will form --- and since this behavior is observable and the event horizon is not, one is led to suggest that this is a candidate for the physical quantity of interest.  

This infinitesimal holonomy has other important features:  it can be shown, in the stationary case, to give the surface gravity; it also makes clear at a quantitative level the way the exponential attenuation and red-shift of signals from matter near the horizon, which was discovered in the early model of \citet{Oppenheimer:1939ue}, extends to more general situations.

These results do need further development.  The situation is cleanest, and best worked out, in the case of radial null geodesics in a spherically symmetric space--time~\citep{Helfer:2001}; there are also results for families of null geodesics approaching the generators of an event horizon, without any symmetry assumptions.  However, for a full connection with potential observables, one also needs to know about families of null geodesics which emerge from near the event horizon, but do not quite approach generators.
Too, one would like to know whether these ideas can help with the teleological problems in classical black-hole thermodynamics.  Finally, precisely because the approach here is so far from teleological, it is not clear how the ideas here are linked to notions of trapping.

\subsection{Spherically Symmetric Black Holes}\label{sphol}

We consider here a spherically symmetric distribution of matter which collapses to form a black hole.  The matter need not be sharply bounded (although it may help to think of this case), but we do require that it fall off, as one moves outwards towards either the future or past along radial null geodesics, that future and past null infinities $\scrif$ and $\scrip$ exist, with their usual Bondi structures.  

We shall use $u$ as the Bondi retarded time parameter at $\scrif$, and similarly $v$ as the Bondi advanced time at $\scrip$.  
Each of these coordinates can be extended uniquely inwards as far as the center of symmetry by requiring them to be spherically symmetric and null.
Note that the spherical symmetry gives us a preferred time direction $t^a$ (orthogonal to the spheres and to $\nabla _a r$).  Both $u$ and $v$ are normalized relative to this, so that $t^a\nabla _au=t^a\nabla _av=1$ (at $\scrif$, $\scrip$).

We define a future-pointing null vector field $l^a$ as follows.  We let $l^a$ be the tangent vector to affinely parameterized radial null geodesics, normalized by 
$l^at_a\Bigr| _\scrif =1$.  (This will, strictly speaking, make $l^a$ ill-defined at the center of symmetry, since any point at the center has a sphere's worth of radial null geodesics through it, but that will not matter.)  Then define a null vector field $n^a$ by requiring that it be future-pointing and directed radially inwards, parallel transported along $l^a$ (so $l^a\nabla _a n^b=0$), and normalized so that $l^an_a=1$.  (It is common to define $l^a$ and $n^a$ like this near $\scrif$, but to use a different choice near $\scrip$.  However, the present system will be most convenient for us.)  Note that $n^a=\partial /\partial u$ and $l_a=du$.

We now define a function which will be of central importance, both here and in the discussion of Hawking's prediction of black-hole radiation.  Imagine following a radial null geodesic backwards in time.  We start from the geodesic's future end-point at some retarded time $u$ on $\scrif$, and end with its past end-point at a retarded time $v=\vv (u)$ on $\scrip$.  The function $\vv (u)$ is, in the sense of geometric optics, the {\em mapping of surfaces of constant phase} for spherically symmetric waves.  (One could also consider, of course, the inverse function $u=\uu (v)$, but most often $\vv (u)$ is what comes up.)

Note that $\vv (u)$ is a strictly monotonically increasing function, and so as $u\to +\infty$ we must have either $\vv (u)\to +\infty$ or $\vv (u)\to v_*$, a finite limit.  the latter case represents formation of a black hole, for then no radial geodesic starting from an advanced time $v\geq v_*$ can emerge to $\scrif$.  The limiting value $v_*$ is {\em the advanced time of formation of the black hole.}

Now consider, in geometric optics, the propagation of a spherically symmetric wave.  If two crests of the wave are separated by a period $\delta v$ at $\scrip$, they will emerge with separation $\delta u$ at $\scrif$, where $\delta v=\vv '(u) \delta u$ (assuming the period is short enough that $\vv '$ is practically constant over the period).  {\em Thus $\vv '(u)$ is the factor by which frequencies are red-shifted,} in the passage of spherically symmetric waves from the distant past to the distant future.   

\begin{wrapfigure}{r}{0.5\textwidth}
\vspace{-2em}
  \begin{flushright}
{\includegraphics[width=.45\textwidth]{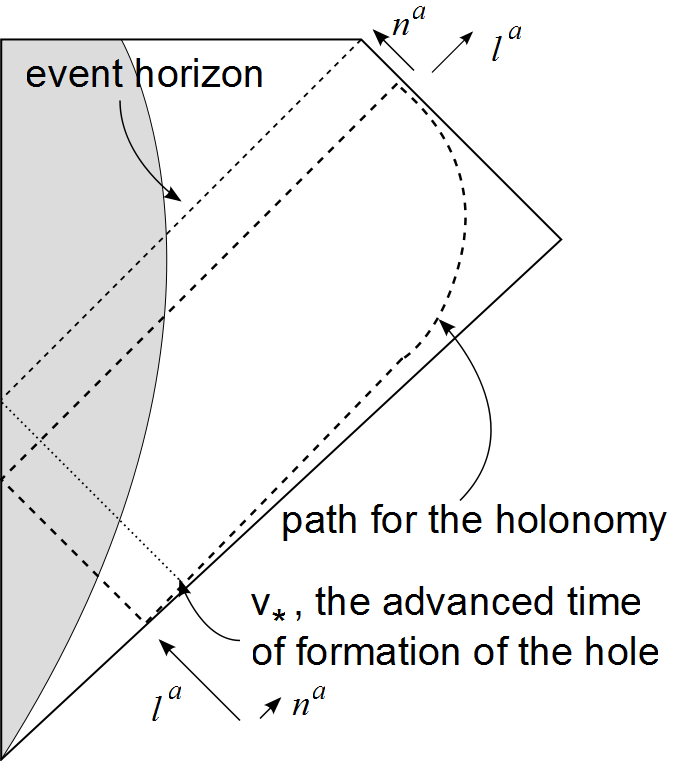}}
\makebox[.45\textwidth]{$\ $ }
\vspace{3em}
\parbox{.45\textwidth}{2.  The path around which the holonomy is taken, in a Penrose diagram.  The shaded region represents the collapsing matter; and $\scrif$, $\scrip$ are at the upper and lower right.}
\end{flushright}
\vspace{-2em}
\label{fig:spherhol}
\end{wrapfigure}

In fact, what we have found is really a holonomy, for we are comparing the parallel propagation of a wave along the radial null geodesic with, implicitly, parallel propagation along a very distant path from the point of emission on $\scrip$ to the point of detection on $\scrif$.  Precisely, if we consider the closed path formed by starting at retarded time $u$ at $\scrif$, going backwards along a very distant path to advanced time $\vv (u)$ at $\scrip$, and then going forward along radial null geodesic to retarded time $u$ at $\scrif$ again, the holonomy acting on vectors orthogonal to the spheres of symmetry is
\begin{equation}
  \Lambda ^a{}_b(u)=(\vv ' (u))^{-1}  l^an_b+ \vv '(u) n^al_b\, .
\end{equation}
Since the curvature gives the infinitesimal parallel transport, we have $\partial _u\Lambda ^a{}_b=-(\int R_{pqc}{}^a l^pn^q\, ds)\Lambda ^c{}_b$,
where we understand that the integrand is 
parallel-transported to the point at the end of the geodesic on $\scrif$.  We thus find that
\begin{equation}\label{vac}
 \vv ''(u)=\vv '(u) \int R_{pqrs}l^pn^ql^rn^s{}\, ds\, .
\end{equation}

Now suppose that the integral $\int R_{abcd}l^an^bl^cn^d\, ds\leq -a<0$.  Then eq.~(\ref{vac}) implies that $\vv '(u)$ will be driven to zero at least exponentially quickly, and we will have $\lim _{u\to +\infty} \vv (u)=v_*$, a finite value.  In this case, an event horizon will form at advanced time $v_*$.  This is in fact typical of what one expects in most spherically symmetric models.  (Conceivably  there could be models in which black holes form ``softly,'' and $\vv '(u)$ decreases to zero but not exponentially quickly.)

There is rather more to the story than this, however.  Suppose that an event horizon will form.  Then $v$ is a good coordinate along the horizon, but $u$ is not, as it diverges.  On the other hand, we can take $U=\vv (u)$ as a coordinate near the horizon; in fact, this coordinate is regular across the horizon (except at the origin), for it is also equal to the advanced time of emission of the radial geodesic which passes outwards through the point.  Thus $U=\vv (u)$ {\em resolves the singularity} $u=+\infty$ at the horizon.  (In the Schwarzschild case, one finds $U$ is a null Kruskal coordinate.)

\begin{wrapfigure}{r}{0.4\textwidth}
\vspace{-1.5em}
  \begin{flushright}
{\includegraphics[width=.35\textwidth]{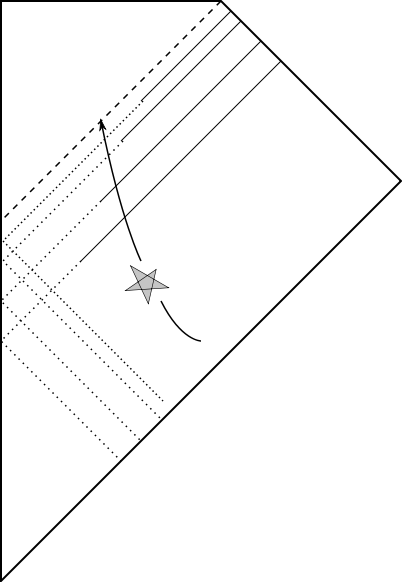}}
\makebox[.35\textwidth]{$\ $ }
\vspace{3em}
\parbox{.35\textwidth}{3.  As later and later geodesics approach $\Hf$, contributions to the holonomy from portions near any given point on $\Hf$ and to the past are suppressed (dotted lines). 
The ratio $-\vv ''/\vv '$ then can be recovered from the logarithmic derivative of the red-shifts of signals from an infalling source,
independent of its peculiar velocity and position.}
\end{flushright}
\vspace{-4em}
\label{fig:spherhola}
\end{wrapfigure}

Now note that we have $n^a=\partial _u=-\vv'(u)\partial _U$.  This the vector $n^a$ will tend to zero as rapidly as $\vv' (u)$ does as we approach any fixed point, or indeed any compact subset, of the horizon.  
On the other hand, the corresponding covector $n_a$ can be viewed as the normalized affine one-form $ds$ along the radial geodesics.  
This means that {\em the affine measure $ds$ along the outgoing radial null geodesics, normalized relative to $\scrif$, tends to zero, as rapidly as $\vv '(u)$ does, along any compact interval
of the geodesic approaching the horizon.}

What this result means is that, not only does $\vv '(u)$ tend to zero because the integral on the right of eq.~(\ref{vac}) is tending to a negative value, but in fact {\em the contributions to the integral from portions of the geodesic near the horizon or to the past of it are suppressed by $\sim \vv' (u)$ as well.}  Thus {\em the ratio $-\vv ''(u) /\vv' (u)$ rapidly becomes independent of all information along the geodesic near, or to the past of, the horizon.}  This ratio can be shown, in the case of a black hole which, once formed, is stationary, to be the surface gravity~\citep{Helfer:2001}.  We thus set $\kappa =-\vv''(u)/\vv' (u)$, the {\em (running) surface gravity} of the incipient black hole.

It is worthwhile emphasizing the limiting process involved here.  We have a family of later and later radial null geodesics, and on these we have segments.  We speak of these segments approaching the horizon in the natural sense of the topological structure of space--time.  This is very different, however, from the 
coordinate-dependent senses of approach often used in the literature.  (Many authors, for example, consider measuring closeness by a Schwarzschild coordinate $r$ or $r_*$, and neglecting $u$.)  

This result has potentially important physical consequences.  Consider some matter falling across the horizon at some event $p$, and emitting radiation radially outwards as it does so.  That radiation emerges at $\scrif$ (in the geometric-optics limit) with a red-shift which is partly due to its climb out of the gravitational potential, but partly due to the peculiar velocity of its source. 
The climb out of the potential in turn depends partly on segments of the null geodesic from the point of emission outwards close to the event horizon, and partly on the climb outwards to the asymptotic regime --- although there is no clean way of saying where one portion of this climb ends and the other begins.  But
what the result above shows is that {\em the logarithmic derivative of the red-shift approaches
$-\vv ''(u)/\vv' (u)$ and so becomes independent of the peculiar velocity of the source, independent of the precise point of emission, and indeed independent of the geometry along any segment of the null geodesic close to any compact portion of the corresponding generator of the horizon},\footnote{The red-shift here is the ratio of the frequencies, that is, the quantity $1+z$ and not just $z$.}  as the horizon is approached.  This is therefore a {\em universal} quantity associated with the horizon:  all matter crossing any compact part of the horizon and emitting signals radially outwards will give rise to the same fractional rate of change of the red-shift factor.

I close this subsection with some comments:  
(a) It ought to be possible to test this universality observationally, and indeed the existence and persistence of it would be circumstantial evidence for a black hole.
(b) It is of considerable interest to study non-radial null geodesics as well as radial ones.
(c) Because of the exponential red-shifts (and associated exponential attentuations), one quickly reaches a point where any real quanta emitted in the vicinity of the horizon have been so severely 
red-shifted on their way out that they are below the limits of any detection.  Thus we are confronted with quantum limitations on the measurement of the vicinity of the incipient black 
hole, a point which will be discussed in section~5.

\subsection{The General Case}

I will now drop the assumption of spherical symmetry, and take up the case of general black holes.  The theory here is less well developed than in the previous case, and I shall indicate the approach, an attractive result, and some problematic issues.

In the previous case, we could use the symmetry to identify a candidate family of null geodesics which would have to approach the generators of a horizon, were one to form:  the radial null geodesics.
In the present, general, case, the problem of non-teleologically determining which escaping null geodesics count as approaching the generators of the horizon (should one form) is more difficult, and a solution to it which is adequate for interpreting 
terrestrial observations will probably require analyzing also 
geodesics which are slightly off-kilter with respect to the 
generators, so that we can say what special properties do 
distinguish those more directly approaching the generators.
Such an analysis has not yet been developed.  Here I will 
suppose that we are following a family of null geodesics which is 
approaching a generator of the horizon, and leave open the 
question of how the family is determined.

Any generator of $\Hf$ is a null geodesic segment; let us consider a one-parameter family $\gamma _u(s)$ of affinely parameterized null geodesics approaching such a segment (along compact subsets), escaping to points $\gamma _u(+\infty )$ at Bondi retarded time $u$ on $\scrif$, and with tangents $l^a$ normalized to the Bondi frame at $\scrif$.
While $u$ and $s$ will not be good coordinates at the horizon, it is nevertheless meaningful to require that the family $\gamma _u(s)$ of geodesics approach the generator smoothly along compact subsets, for example by requiring that there are two curves $s=s_1(u)$, $s=s_2(u)$ for which $\gamma _u(s_j(u))$ tend smoothly to two distinct points on the generator.
Note that there will be many one-parameter families of null geodesics approaching the same generator.

We will suppose that the vector field $l^a$ along this 
one-parameter family of geodesics has been completed to a
parallel-propagated standard null basis ($l^a$, $m^a$, ${\overline m}^a$, $n^a$, the only 
non-zero inner products among them being $l^an_a=1$ and $m^a{\overline m}_a=-1$).  Then $w^a=\partial _u\gamma$ is a connecting vector field, satisfying the Jacobi equation
\begin{equation}
  (l\cdot\nabla )^2w^a=l^pl^qR_{pbq}{}^aw^b\, .
\end{equation}
Using the peeling of the curvature in the Bondi--Sachs asymptotics, one can show that $w^a\sim n^a +s(\omega  m^a +\overline\omega {\overline m}^a)$ (plus a possible multiple of $l^a$) for some complex $\omega$ asymptotically, and we may identify 
$\omega  m^a +\overline\omega  {\overline m}^a$ as the (running, with respect to $u$) angular velocity of the
family of geodesics approaching the particular generator of $\Hf$, relative to $\scrif$.  (In the present, fully dynamical, situation, it is not clear that one can define a ``running angular velocity'' of a generator on $\Hf$, independent of the family used to approximate it.)

Let us now consider the construction of a holonomy operator.  
Without additional assumptions, there is no guarantee that the extensions of the geodesics $\gamma _u(s)$ to the past reach $\scrip$.  On the other hand, we learned from the spherically symmetric case that for many purposes it is not necessary to consider the details of the pasts of these geodesics.  Let us rather fix any point $p$ of interest on the generator of $\Hf$ that the geodesics are approaching, determined by $\lim _{u\to +\infty}\gamma _u(s(u))$ for some function $s(u)$.  
Consider the holonomy along a path from the 
end-point $\gamma _u(+\infty )$ of one of the geodesics, backwards along $\scrif$ to $\gamma _{u_0}(+\infty )$, then backwards along the $u_0$-geodesic to $\gamma _{u_0}(s(u_0))$, then along $\gamma _u(s(u))$, and finally outwards along the $u$-geodesic again.

In this non-symmetric case, the infinitesimal holonomy cannot be characterized simply by a scalar, but will be a generator of Lorentz motions.  It is again given by integrating the curvature; the contribution from the geodesic\footnote{If there is gravitational radiation present, it contributes interesting effects to the portion of the path along $\scrif$.  However, these do not affect the main points and will be discussed elsewhere.}
is
\begin{equation}
  \lambda ^a{}_b=-\int R_{pqb}{}^al^p w^q\, ds\Bigr| _\Hf ^\scrif\, ,
\end{equation}
where we understand that the integrand has been parallel-propagated along $\gamma _u$ to $\gamma _u(+\infty )$, and the limits of integration stand for the point $\gamma _u(s(u))$ near $\Hf$ and the point $\gamma _u(+\infty )$ on $\scrif$.
We note the identity 
\begin{equation}\label{lambdident}
  \lambda ^a{}_b l^b=-l\cdot\nabla w^a\Bigr| _\Hf ^\scrif\, ,
\end{equation}
where
we understand that the vector $l\cdot\nabla w^a$ at
the lower limit must be parallel-transported to the point at $\scrif$.

Let us now consider the case of a black hole which becomes stationary, and let $w^a=\chi ^a$ be the
Killing field which becomes tangent to the generators of the horizon, as usual.  If we contract both sides of eq.~(\ref{lambdident}) with $w^a(\Hf )$, we get
\begin{equation}\label{sugra}
 \lambda ^a{}_bl^bw_a(\Hf )=-w_a(\Hf )(l\cdot \nabla w^a )(\scrif ) 
  +w_a(\Hf )(l\cdot \nabla w^a)(\Hf )\, .
\end{equation}
In the limit of later and later geodesics, the first term on the right will vanish in the stationary case, for $l\cdot\nabla w^a (\scrif ) =\omega m^a+\overline\omega {\overline m}^a$, and $w^a(\Hf )$ will become proportional to $l^a$.  In this limit, the second term (divided by $l\cdot w(\Hf )$) becomes a well-known expression for the surface gravity, and thus we conclude that, in the stationary case, the limiting value of $ \lambda ^a{}_bl^bw_a(\Hf ) /l^aw_a(\Hf )$ represents the surface gravity, where $w^a$ is chosen to be the usual Killing field.  We have therefore
\begin{equation}
\kappa =\lambda ^a{}_bl^bw_a(\Hf )/l^aw_a(\Hf )\, ,
\end{equation}
where we understand the symbol $\Hf$ is a short-hand for taking the limit as the event horizon is approached.
We therefore get a new interpretation of surface gravity, in case of stationary black holes, as this infinitesimal holonomy.  The quantity~(\ref{sugra}) is well-defined even in the non-stationary case, however, and may be of interest to those searching for a general definition of surface gravity.

\section{Quantum Fields in Curved Space--Time}

While we are ultimately interested in those aspects of quantum field theory  related to black holes, it will be helpful in this section to first review the general theory in curved space--time in the absence of extreme circumstances. 

The present theory of quantum fields in curved space--time is so natural and attractive that one should be chary of trying to modify it.  Still, it would be wise to keep the following points in mind:

\begin{itemize}

\item
It is almost entirely untested experimentally.  While we have evidence for many
quantum-field-theoretic processes operating in the Universe, none of these so far probes deeply the distinctive elements of the curved-space theory.

\item
While the theory does give an extremely natural prescription for computing some important observables (for instance, field-strength operators and particle number density operators asymptotically far from sources), it
runs into difficulties when we try to push it past a certain point, especially when we try to construct the stress--energy operator.  
\end{itemize}

\noindent This latter point is not usually considered to be very important:  a fly in the ointment only --- but I shall explain
in Section 5 why I believe it may be a hint of more serious foundational concerns:  a cloud on the horizon.

A few remarks for students:  The development of quantum field theory took decades, both to work out techniques of computation and to sort out interpretational issues, and there are still some foundational problems with it, while at the same time it is unarguably powerful and successful.  Some parts of the field are understood in a practical sense better than they are in a foundational one.  This is only to be expected in such a conceptually difficult area, and is a mark of how close the material is to the frontier of research.  Expect to learn the material in stages and to refine your understanding as you go along --- and always try to verify claims you encounter.

\subsection{Contrast with Special-Relativistic Theory}

The development of quantum field theory in curved space--time forced physicists to reconsider the foundations of the usual theory.  In doing so, it became apparent that much of the
way the special-relativistic theory is usually explained --- starting with Fock spaces and creation and annihilation operators --- relies heavily on intuitions about how the theory should come out.  These intuitions turn out to be correct in Minkowski space, but the difficulty is that appealing to them skips over the issues of just how they are justified --- and in curved space--time, where the intuitions are not generally correct, we need to look at what the justifications were and how the conclusions must be modified.

The most fundamental point is that quantum field theory is, primitively, a theory of {\em fields,} and {\em particles are a derived concept} in it.  This is of course almost backwards of the way the special-relativistic theory was developed, is usually taught, and is generally used by particle theorists.    Nevertheless, to understand how the theory is generalized to curved space--time, this is the key point.  We must think, for example, of quantum electromagnetism as being, at a primitive level, about the electric and magnetic field strength operators (or more properly their averages over small space--time volumes), rather than about photons.  (Willis Lamb wanted to issue ``photon licenses'' only to those people who could use the term properly.  He presumably did not have curved 
space--time in mind, but he had a deep sense of how less-than-straightforward the particle concept was.)

In fact, particles (and the related concepts of the vacuum --- the no-particle state --- and ladder operators) depend heavily on Poincar\'e invariance.  They are defined by resolving the fields into positive- and negative-frequency parts, and it is a critical consequence of Poincar\'e invariance that the same positive-/negative-frequency decomposition is obtained no matter which inertial time axis is used for the Fourier transformation.  However, in curved space--time, such a decomposition would indeed depend on the choice of timelike curves one Fourier-analyzed along.

One might think this is caviling, that in familiar cases the space--time curvature is such a tiny effect that that there is little substantive ambiguity in the definitions of positive and negative frequency.  And indeed this is true for ordinary particle-physics purposes.  But in crucial cases --- in particular, for Hawking's work on black holes --- it is just these ambiguities, and their physically correct resolution, which are at the heart of things.

\subsection{Linear Fields in Minkowski Space}

While the case of curved space--time differs from that of Minkowski space, it is worthwhile sketching the special-relativistic case here.  In particular, I want to emphasize some points not often discussed in introductory treatments but important for us:  the nonlocality of particle concepts, and the character of quantum field fluctuations.

In order to have a manifestly relativistically invariant theory, we work in the Heisenberg picture, so that the field operators will be functions on space--time (more properly, they will be operator-valued distributions).  The state vector will be fixed, except when a quantum measurement is made, when it will be projected into a subspace according to the usual rules of quantum theory.  However, those comments describe what we are aiming at;  we must show how to achieve it:  to construct the operators and the Hilbert space.

Let us consider a massless (there is little difference in the massive case) real scalar field satisfying the wave equation
\begin{equation}
  \nabla _a\nabla ^a\phi =0
\end{equation}
and the canonical commutation relations
\begin{eqnarray} 
 [\partial _t\phi (t,{\bf x}),\phi (t,{\bf y})] &=&i\delta ^{(3)}({\bf x}-{\bf y})\\
 \left[\phi (t,{\bf x}),\phi (t,{\bf y})\right]&=&
 [\partial _t\phi (t,{\bf x}),\partial _t\phi (t,{\bf y})]=0 \, .
\end{eqnarray}
To quantize this, one typically proceeds as follows:

\begin{enumerate}

\item One Fourier-transforms the field, using the field equations to write
\begin{equation}
  \phi (t,{\bf x})=\int \left( e^{iEt-i{\bf k}\cdot {\bf x}} a({\bf k}) +e^{-iEt+i{\bf k}\cdot {\bf x}} a^*({\bf k})\right) 
  \frac{d^3{\bf k}}{2^{1/2}(2\pi )^{3/2}E^{1/2}}\, ,
\end{equation}
where $E=E({\bf k})=\| {\bf k}\|$.  At this point, the q-numbers $a({\bf k})$, $a^*({\bf k})$ have yet to be determined.

\item One then examines what the canonical commutation relations imply about $a({\bf k})$, $a^*({\bf k})$, recognizing in them the usual quantum-mechanical {\em ladder algebras}:
\begin{eqnarray}
  [a({\bf k}),a^*({\bf l})] &=& \delta ^{(3)}({\bf k}-{\bf l})\\
  \left[a ({\bf k}),a({\bf l})\right] &=&
      [a^* ({\bf k}),a^*({\bf l})]=0\, .
\end{eqnarray}
Notice that at this point we have not yet constructed the Hilbert space, so it would not really be correct, at this stage, to refer to the fields or the q-numbers as {\em operators} (they do not yet operate on anything).  What we have defined at this point is the {\em field algebra} $\A$, that is, the commutation relations the fields must satisfy, rather as if we had defined a group by a multiplication table but had not yet given a representation of it by matrices acting on a vector space.

\item One then constructs the {\em Fock representation}, by {\em assuming that there exists a Poincar\'e-invariant ground state} $|0\rangle$ which is normalized and annihilated by all the $a({\bf k})$.
The rest of the construction follows directly by applying the canonical commutation relations, just as one shows in elementary quantum mechanics that one can deduce the spectrum of the harmonic oscillator from the ladder-operator algebra together with the assumption that there is a normalizable ground state.

\item Finally, one establishes the particulate interpretation of the theory by showing 
that the vacuum and one-particle states are the appropriate eigenstates of the mass and spin operators, and the $n$-particle states are tensor products of the one-particle states.

\end{enumerate}

There are a number of important points to make about this:

(a) The entire procedure uses Poincar\'e invariance, and most especially time-translation invariance, very strongly.  It is this which gives us the split of the field operators into positive- and negative-frequency parts, and the vacuum state, from which all others are constructed, is characterized in terms of this split.

(b) There are other representations of the field algebra, but they do not admit 
Poincar\'e-invariant vacuum states.

(c)
The particle states are non-local.  If, for example, we work out the inner product on one-particle states, we find that it has the form
\begin{equation}
  \langle \phi |\psi\rangle =\int {\overline\phi}(0,{\bf x})K({\bf x},{\bf y})\psi (0,{\bf y})\, d^3{\bf x}\, d^3{\bf y}\, ,
\end{equation}
where the kernel $K({\bf x},{\bf y})$ is not simply the delta function $\delta ^{(3)}({\bf x}-{\bf y})$, but is non-zero even for ${\bf x}\not={\bf y}$.  This non-locality is not very important for modes of wavelengths above the Compton scale $\|{\bf k}\| ^{-1}$.\footnote{Even in the massive case, the inner product is nonlocal on the scale $(\| {\bf k}\|^2+m^2)^{-1/2}c$.  In the non-relativistic case, this reduces to the Compton wavelength, which is very small.}  However, for 
shorter-wavelength modes the non-local nature of the kernel must be taken into account.
Note that this means {\em it is impossible to say that a particle is within a volume of space of characteristic size $\lesssim \hbar c/E({\bf k})$.}

(d) Measurements of the field operators in arbitrarily small space--time volumes are defined, however.  But the particle states, and in particular the vacuum state, are {\em not} eigenstates of these operators.  For example, if $\Phi (a) =(4\pi a^3/3)^{-1}\int _{{\bf x}\leq a}\phi (0,{\bf x})\, d^3{\bf x}$ is the average of the field over a sphere of radius $a$, then one finds $\langle 0|\Phi (a)|0\rangle =0$ but
\begin{equation}
\langle 0|(\Phi (a))^2|0\rangle\sim a^{-2}\mbox{  as  }a\to 0\, ,
\end{equation}
the larger and larger fluctuations as $a\to 0$ representing the deviation of the vacuum state from an eigenstate of the field operator.  These are the famous {\em vacuum fluctuations}.
It will be vacuum fluctuations which will ultimately be responsible for the production of quanta in Hawking's proposal of black-hole radiation.

(e) The two-point function
\begin{equation}\label{twopoint}
\langle 0|\phi (p)\phi (q)|0\rangle =
  -\frac{1}{4\pi ^2}\cdot\frac{1}{(p-q-i\epsilon )^2}
     \, ,
\end{equation}
where I am now usuing $p$, $q$ to stand for points in Minkowski space and $\epsilon$ is an infinitesimal future-directed timelike vector, plays a key role.\footnote{Often one works with the time-ordered two-point function, but this is unnecessary here.} In fact, for any ``reasonable'' state $|\Psi\rangle$, the two-point function $\langle\Psi |\phi (p)\phi (q)|\Psi\rangle$ will have the same ultraviolet asymptotics (that is, as $p\to q$), since there will be an energy scale beyond which the state is unexcited and on correspondingly small space--time scales the state will ``look like'' the vacuum.

(f) Note in particular that the two-point function 
does not vanish at spacelike separations.  This means that {\em the vacuum state contains field correlations at spacelike separations.}

A consequence of this is
that one cannot simply restrict the Fock quantization to a subvolume of Minkowksi space, for the fields in that volume are correlated with those elsewhere.  For many problems, of course, the effects of such correlations are negligible; on the other had, in some cases they are important and one must look to the precise physics of the situation.  
But it is the contrapositive of this which is the most important for many of the arguments in this area:  it is generally impossible to ``quantize in a given volume''  and have the results there accurately reflect the Fock quantization.  In particular, arguments about ``quantization in the Rindler wedge'' should be examined carefully.

The {\em origin} of the spacelike correlations in the field is a profound, ultimately cosmological, problem.

\subsection{Construction of a Quantum Field Theory}

I shall here outline the construction of the theory of a linear, massless, minimally coupled real scalar field in curved space--time. 
This is really a stand-in for the physically more interesting case of the electromagnetic field, but the constructions depend little on the mass or spin.  The main advantage of the scalar case is that we do not have to deal with technical issues relating to the quantum treatment of gauge invariance; such issues do not affect the points we will make.
As above, we will work in the Heisenberg picture.

It is helpful to think of the construction in two stages.  First, we construct the {\em field algebra} $\A$, that is, we specify the commutation relations which must hold.  One should think of this as an abstract algebraic structure, much like specifying a group by giving its multiplication table.  The second stage is to construct a {\em representation} of that algebra, that is, a physical Hilbert space $\HH$, together with an interpretation of the fields as
operators on that space, satisfying the correct commutation relations.
In order to have a sensible theory, we require the space--time to be globally hyperbolic, that is, it admits Cauchy surfaces on which initial data for the field may be given resulting in a
 well-posed evolution problem. 

Let us denote the field $\phi$.  Then the field algebra $\A$ is characterized by two requirements:  that the field equation
\begin{equation}
  \nabla ^a\nabla _a\phi =0
\end{equation}
be satisfied; and that the canonical commutation relations
\begin{equation}
  [t^a\nabla _a\phi (p),\phi (q)] =i\delta _\Sigma (p,q)\, ,\quad
 [\phi (p),\phi (q)]=[t^a\nabla _a\phi (p),t^b\nabla _b\phi (q)]=0
\end{equation}
on any Cauchy surface $\Sigma$ with
future-directed normal $t^a$, hold.\footnote{The formal definition is based on this idea but phrased somewhat differently for technical reasons.}  In fact, the canonical commutation relations can be written in a manifestly invariant form first noted by Peierls:
\begin{equation}
  [\phi (p),\phi (q)] = \Delta _{\rm a}(p,q) -\Delta _{\rm r}(p,q)\, ,
\end{equation}
where $\Delta _{\rm a}$, $\Delta _{\rm r}$ are the advanced and retarded Green's functions for the field equation.

We now turn to the representation.

In quantum mechanics, the Stone--von Neumann Theorem asserts that all representations of the canonical commutation relations are unitarily equivalent.  Thus in quantum mechanics there is no question of principle as to which representation to use:  one may be more convenient than another for some purposes, but all are interconvertible.
In quantum field theory, the situation is different.  There, there are an infinite number of distinct equivalence classes of representations, and one must decide which are physically appropriate.  In the special-relativistic theory, this was done by requiring the existence of a Poincar\'e-invariant state, the vacuum; here, we must find another criterion.

It turns out that the equivalence classes of the representations are essentially specified by the asymptotic forms of the two-point functions
\begin{equation}
  \langle\Psi | \phi (p)\phi (q) |\Psi\rangle
\end{equation}
for $p$ and $q$ close to each other, and as one (or both) recede to infinity.
These correspond, in standard parlance, to the ultraviolet and infrared asymptotics of the theory.  We shall not worry about infrared issues, which typically bear on only cosmological questions.  On the other hand, the ultraviolet asymptotics go to the question of how the theory, on short scales, compares with that in Minkowski space.\footnote{The precise state used in this relation turns out not to matter very much, because in the ultraviolet limit one expects all of the modes to be unoccupied.}

It turns out that there is a distinguished equivalence class of representations (modulo infrared issues) for which the leading ultraviolet asymptotics have the same form as in Minkowski space.  These are the {\em Hadamard} representations, and they are the physically natural candidates for quantization.  It is an important and non-trivial theorem that these are well-defined, that is, that the correct ultraviolet asymptotics are preserved under evolution.

We assume henceforth that we have constructed a Hadamard representation.  We can do this in practice, for instance, by selecting a Cauchy surface and identifying the Cauchy data with those for the corresponding field equation on a Cauchy surface in Minkowski space in the usual Fock representation, and then defining the field elsewhere by its evolution.  
  
We now have, in principle at least, enough data to define the quantum field theory.  We have shown that there is a well-defined field algebra $\A$, and a preferred (up to infrared issues) representation of that algebra as operators on a Hilbert space $\HH$.  Of course, the description is so far fairly abstract, and we must develop it in more detail if we are to learn how to analyze particular physical problems.
There are two main issues to  take up:  how to effectively compute the evolution of the fields; and how to specify (and analyze) the physical content of the states.  The former depends on the particular space--time, but the latter is a more general issue.

It has sometimes been suggested that one can dispense with the construction of representations in the sense I have described, and
work only with the algebra $\A$ and what can be thought of as generalized density matrices (and it is these which are called states in the algebraic approach).  While this is adequate for some purposes, any density-matrix-like approach does not detect phase information which is generally important in quantum theory; also, one must somehow cut the freedom in the choice of class of admissible density matrices down to correspond to that in an irreducible representation of the field algebra, and in doing this one must either use the Hadamard condition or come up with a physically plausible substitute.

\subsection{The Physical Content of a State}

I emphasized earlier that the primitive elements of our theory are the quantum fields.  I have been discussing, for simplicity, a scalar field $\phi$, for which the field measurements would correspond to (weighted averages of)  $\phi $ and its canonically conjugate momentum $t^a\nabla _a\phi$ near a hypersurface with normal $t^a$, but this is really a stand-in for the more physically realistic case of the electromagnetic field, where one measures (weighted averages of) the  magnetic ${\bf B}$ and electric ${\bf E}$ field strength operators.\footnote{Because only gauge-invariant quantities are observable, one cannot directly measure the potential.}  These operators ($\phi$ and $t^a\nabla _a\phi$, or ${\bf B}$ and ${\bf E}$) are 
complementary, and so one cannot measure them simultaneously.  Thus one could specify the state by giving a ``wave functional'' $\Psi _q[\varphi ]$ or $\Psi _q [{\mathcal B}]$ analogous to the position wave function in quantum mechanics, or a momentum-type functional 
$\Psi _p[t^a\nabla _a\varphi ]$ or $\Psi _p[{\mathcal E}]$, in each case referring to the field eigenvalues on a specific hypersurface.  These wave functionals are just the coefficients of the corresponding eigenstates of the field measurements, just as the wave functions in quantum mechanics are the coefficients of the corresponding position or momentum eigenstates.\footnote{The mathematical technicalities in treating these wave functionals are more involved than ordinary quantum mechanics, owing to the infinite-dimensionality of their domains.}  Such expressions are said to concern the {\em field aspect} of the state.

At least in special-relativistic theory, however, we are more often concerned with a state's {\em particle aspect} than its field aspect, that is, we wish to know its particle content.  How is this defined?

In special relativity, the particles are supposed to be eigenstates of mass, spin, and other appropriate quantum numbers.  The mass and spin (or, more precisely, their squares) are in fact the Casimir operators for the Poincar\'e group, and so analyzing the particle-content of a state involves decomposing it into irreducible representations of the Poincar\'e group.

In the case of linear quantum fields, the end-result of this is characterized by {\em number density operators} associated with the different modes of the field, analogous to the number operator for the harmonic oscillator.  For the scalar field discussed earlier, these would be 
\begin{equation}
  n({\bf k})=a^*({\bf k}) a({\bf k})\, ;
\end{equation}
the number density of particles with momentum $\hbar{\bf k}$.
(For the electromagnetic field, the operator depends on the polarization as well as the wave-number ${\bf k}$.)
One sees quite explicitly that this depends on the Fourier transform of the field, which in turn depends heavily on the Poincar\'e invariance.

Poincar\'e invariance need not hold exactly for us to have a reasonably good, although in principle imperfect, definition of number density, however.  Evidently, if we have a region of space--time which is well-approximated by a portion of Minkowski space of linear size $\sim l$, then the Fourier components of wave-packets within that region, of nominal wave-numbers $\gg l^{-1}$, will be well-defined, and we will have good approximate number-density operators for the corresponding modes.  In fact, since all real particle detectors function only over finite volumes of space--time, they really respond to such approximate number density operators only.  The discrepancy between these and the true operators is analogous to considering the corrections to an optical system due to finite-aperture and finite integration-time effects.

In many cases one can identify regions far from the gravitational sources in which the curvature is quite small and one thus has good definitions of number density operators for at least substantial regimes in wave-number space.  In fact, often we can cover hypersurfaces in the distant past or future with regions which are approximately Minkowskian, and then we can analyze the particle-contents within these regions fairly well, and get overall characterizations of the states up to the problems associated with patching the regions together.
On the other hand, when we attempt to link the vicinity of a black hole's event horizon to the region near $\scrif$, we have seen that the holonomy typically involves exponentially growing an decaying terms, and thus curvature effects over this entire regime cannot be discounted.  We will examine this in more detail in the next section.

The reader may have noticed that something curious has crept in in the localization of the particle content just discussed:  while we may have good (approximate) definitions of particle number-density at different times, there is no guarantee that those at one time should be compatible with those at another.  This is indeed the case, and corresponds to the possibility of particle creation, as will now be discussed.

\subsection{Bogoliubov Transformations}\label{BTsec}

Suppose that we have a field $\phi$ propagating through space--time, and on two Cauchy hypersurfaces $\Sigma _{\rm p}$, $\Sigma _{\rm f}$ we have physically-justified resolutions of the field into positive- and negative-frequency modes and associated annihilation and creation operators $a_{\rm p,f}({\bf k})$, $a^*_{\rm p,f}({\bf k})$.  Here, the subscripts ``p,'' ``f'' are meant to suggest past and future, but that is to fix ideas only.  The relative positions of $\Sigma _{\rm p}$, $\Sigma _{\rm f}$ are immaterial, and they may overlap.  Also, I have written the mode label ${\bf k}$ as if it were a wavenumber, but this interpretation is not necessary; one only needs the ladder algebra to be obeyed.
A simple if overidealized example would occur in a space--time which was initially stationary and close to Minkowskian, then passed through a time-dependent phase, and returned finally to a stationary, approximately Minkowskian, state.

Because the field algebra can be expressed equally well on either Cauchy surface, we may express the annihilation operators with respect to $\Sigma _{\rm p}$ in terms of those at $\Sigma _{\rm f}$ (or vice versa):
\begin{equation}\label{bog}
  a_{\rm p}({\bf k}) =\int \alpha ({\bf k},{\bf l}) a_{\rm f}({\bf l})d^3{\bf l}
  +\int \beta ({\bf k},{\bf l}) a^*_{\rm f}({\bf l})d^3{\bf l}\, ,
\end{equation}
or more briefly
\begin{equation}\label{bogb}
  a_{\rm p} =\alpha  a_{\rm f}
  +\beta a^*_{\rm f}\, .
\end{equation}
That is, because the field has propagated through a time-dependent region, its modes have mixed relative to the two different ways of resolving it.  The quantities $\alpha$ and $\beta$ are called {\em Bogolioubov coefficents}, and eqs.~(\ref{bog}), (\ref{bogb}) are called a {\em Bogoliubov transformation}.  Note that the field modes, that is, the coefficients of the annihilation and creation operators in the expansion of the field operators, will also transform via a Bogoliubov transformation, contragredient to the q-numbers $a$, $a^*$.

We may construct Fock-like representations of the field relative to each of $\Sigma _{\rm p,f}$, beginning in each case with a  vacuum state $|0_{\rm p,f}\rangle$ characterized by $a_{\rm p,f}|0_{\rm p,f}\rangle =0$, and applying the creation $a^*_{\rm p,f}$ and annihilation $a_{\rm p,f}$ operators.  These will both be Hadamard representations (assuming that the positive-/negative-frequency decompositions at $\Sigma _{\rm p,f}$ were physically correct, that is, gave the right ultraviolet asymptotics), and so they will be equivalent, but there will be a non-trivial transformation between them.

We can see this explicitly.  Suppose the state is the p-vacuum $|0_{\rm p}\rangle$, characterized by $a_{\rm p}|0_{\rm p}\rangle =0$.  Using eq.~(\ref{bogb}), we can rewrite this condition as
\begin{equation}
 (\alpha  a_{\rm f}
  +\beta a^*_{\rm f})|0_{\rm p}\rangle =0\, .
\end{equation}
We can regard this as an equation for $|0_{\rm p}\rangle$ in terms of the data at $\Sigma _{\rm f}$.  In fact, simply using the ladder algebra, we find
\begin{eqnarray}\label{vactr}
  |0_{\rm p}\rangle &=&\mbox{(normalization)}\cdot
      \exp [-(\alpha ^{-1}\beta /2)a^*_{\rm f}a^*_{\rm f}]
   \, |0_{\rm f}\rangle\\
&=&\mbox{(normalization)}\cdot 
  \sum _{n=0}^\infty \frac{(-(\alpha ^{-1}\beta /2)a^*_{\rm f}a^*_{\rm f})^n}{n!}\, 
  |0_{\rm f}\rangle \, ,
\end{eqnarray}
showing that the p-vacuum appears as a superposition of states of different f-particle numbers, the first contribution being an f-vacuum one, then a two f-particle one, and so on.  It is sometimes said that eq.~(\ref{vactr}) shows that the p-vacuum is a Gaussian distribution of 
f-particles, but one should bear in mind that this equation really applies at the level of probability {\em amplitudes} and not probability {\em distributions}.

Formulas like~(\ref{vactr}) show that {\em the particle content of a state is a function of where it is observed}, and is not absolute.  
There is a sort of limited observer-independence, in that it may well happen that in regions around $\Sigma _{\rm p}$ or $\Sigma _{\rm f}$ separately there may be agreement among relatively boosted, rotated or translated observers about the particle-content, but there will not be agreement in comparing $\Sigma _{\rm p}$ with $\Sigma _{\rm f}$.
The physical interpretation of this is that the passage of the field through a time-dependent potential, or region of time-dependent curvature, has resulted in the creation or destruction of particles.  Just this point was a main reason for developing quantum field theory in contradistinction to quantum mechanics:  the need to accommodate changing particle numbers.

A well-known example of this --- predicted, but not yet experimentally verified --- is the {\em Schwinger effect}, where one considers the electron--positron field propagating through a time-dependent classical electromagnetic potential~\citep{Schwinger:1951nm}.
Even an initially vacuum state of the charged field is predicted to give rise to pairs of real charged particles as time passes.

It is important to note that one consequence of this observer-dependence of particle-content is that there will be regimes in which there is no very good definition of particle-content.  For instance, in a region where the potential or the space--time geometry is changing on a timescale $\sim T$, it will not be possible to give much meaning to the notion of particles with energy $\sim\hbar /T$.  
This issue affects all of the arguments which attempt to explain black-hole radiation quantum-{\em mechanically}, that is, without confronting the {\em field-theoretic} issues bound up with the ambiguity in the definition of particles.

\subsection{The Unruh Process}

\citet{Unruh:1976db}, in studying the foundations of Hawking's black-hole radiation prediction, considered the response of a detector uniformly accelerating in Minkowski space, and predicted that it would respond as if it were in a thermal bath of temperature $T_{\rm U}=\hbar a/2\pi c$ (with $a$ the acceleration, Boltzmann's constant being taken to be unity).  This is not really a prediction of quantum field theory on curved space--time --- the production of the quanta coming from the acceleration of the detector rather than any space--time curvature or even potential --- but it is closely allied to those ideas.  And as it is often linked with Hawking's prediction, it is worth outlining here.

Consider a scalar field $\phi$ in Minkowski space, and suppose there is a measuring device which follows a world-line $\gamma (s)$ parameterized by proper time.  We will be mostly interested in the case where the world-line is initially inertial, and then smoothly changes to uniformly accelerate for a period.  Because we shall only be interested in causal considerations, the behavior of the world-line as $s\to +\infty$ will not enter.

I have so far not specified just how the detector responds, and indeed this will depend on just how it is constructed.  But the main point is that the detector can only respond to the field $\phi (\gamma (s))$, since it is local to the world-line.  Thus it cannot really be a particle detector, for that would require averaging over some finite spatial extent, a feature we are (so far) ignoring; it is a field-strength detector, or more precisely, it responds to weighted averages of field strengths along the world-line $\gamma (s)$.
Thus the portion of the field algebra accessible to the detector is that generated by $\phi (\gamma (s))$.  Whatever the internal physics of the detector is which determines what it measures, the assumption is that it will respond according to its internal, proper, time.

It should be evident that the physics of the operators $\phi (\gamma (s))$ will depend on the particular world-line $\gamma (s)$.  We can see this already in considering the two-point function
$\langle 0|\phi (\gamma (s_1))\phi (\gamma (s_2))|0\rangle$.
This function can be interpreted as describing the spectrum of vacuum fluctuations of field measurements relative to the proper time along the world-line, and that spectrum will depend on the details of world-line.   Thus the results of measurements by an accelerating detector will in general depend on the details of the acceleration.

In particular, if for some range of $s$-values the world-line corresponds to uniform acceleration, say
\begin{equation}
  \gamma (s) =(a^{-1}\sinh (as),a^{-1}\cosh (as),0,0) 
\end{equation}
in a Cartesian coordinate system, then one has
\begin{equation}
\langle 0|\phi (\gamma (s_1))\phi (\gamma (s_2))|0\rangle =
 -\frac{1}{4\pi^2}\frac{a^2}{(\sinh as_1-\sinh as_2 -i\epsilon )^2-(\cosh as_2-\cosh as_1)^2}\, .
\end{equation}
Note that this is periodic in imaginary time with angular frequency $T_{\rm U}=\hbar a/(2\pi c)$.  This sort of periodicity is characteristic of thermal behavior, and indeed one can check that the $n$-point functions are those of a thermal field at this Unruh temperature, over the interval in which the world-line accelerates uniformly.

A few comments about this are in order:

Unruh's analysis is sometimes criticized on the grounds that particle states have finite spatial extent, and he worked only along a given world-line.  However, the treatment I have given avoids this problem, by using not particle, but field-strength, detectors.  
It has also been criticized because he considered a world-line accelerating for all time.  The treatment here avoids this problem too.
I believe this analysis is secure, but there has been no experimental verification of Unruh radiation as yet.

Many authors have noted that a uniformly accelerated detector follows an integral curve of a boost Killing vector field in Minkowski space, and have considered the ``Rindler wedge'' formed by a connected family of such curves which are timelike.  There is certainly much beautiful geometry in such analyses, but some cautions should be borne in mind.  First, one cannot truly ``quantize in the Rindler wedge'' alone and have the theory really represent a restriction of a full Minkowski-space theory to the wedge, because, as we have seen, the field in the wedge is correlated with the field outside of it.  Thus, if the relevant space--time is really Minkowski space, any work purely in the Rindler wedge must be justified in terms of the theory in the full Minkowski space.  And finally, the mathematical structure of the Rindler wedge is a result of the very high degree of symmetry present, and it is hard to draw general lessons from such special situations.

Let me close by mentioning a very beautiful result which was obtained at about the same time Unruh's was, in axiomatic field theory.  The Bisognano--Wichmann theorem \citep{Bisognano:1976za} asserts that in a Poincar\'e-invariant theory with non-negative energy spectrum, even an interacting theory, the $n$-point functions of the fields along any uniformly accelerating world-line are the same as those of a non-accelerating observer but seeing the theory in a thermal (KMS) state at the Unruh temperature.

\section{Hawking's Prediction of Black-Hole Radiation}

In 1974 and 1975, Hawking asserted that, taking quantum field theory into account, black holes are not in fact black, but emit thermal radiation~\citep{Hawking:1974rv,Hawking:1974sw}.  Whether this prediction turns out to be correct or not, Hawking's work contributed immensely to advancing our understanding of quantum fields in curved space--time.  
That there are problems with the analysis (and with others which have been offered in support of its conclusions) is because
physical issues still deeper than those originally considered are implicated.

The issues are fundamental.  The point not so much that there is a question about whether black holes radiate, as that Hawking's work leads directly to the conclusion that the theory of quantum fields in curved space--time is {\em inadequate} for treating the system. 
The core of his analysis is a very beautiful and insightful 
understanding of the propagation of the field modes through the black-hole space--time,
but just that analysis shows 
that the Hawking quanta have, as their precursors in the distant past, vacuum fluctuations of exponentially increasing frequencies.  Those frequencies quickly pass the Planck scale (and indeed any scale, including the total estimated energy of the Universe), and quantum field theory surely becomes inadequate.  This is the {\em trans-Planckian problem}.

The force of this problem has not always been appreciated, for two reasons.  The first is that its invariance has not always been recognized; this has led a number of workers to suggest that it is an artifact of the way Hawking's computations were done and might be avoided by suitable mathematical transformations.
But the problem is invariant, and 
we shall see that the holonomy ideas discussed earlier clarify this issue.  

The second reason that the severity of the trans-Planckian problem has not always been appreciated is that it is a {\em virtual} issue.  It does not appear explicitly in the predicted thermal radiation, and thus the failure of the theory is not apparent.  Indeed, at the time Hawking wrote his papers, the significance of a virtual trans-Planckian problem was clear only to a small number of workers.  However, this point is more broadly understood now.  And in Section~\ref{moreqft} I shall show that this problem is in any case promoted from virtual to real when interactions or measurement effects are taken into account.

Closely related to the trans-Planckian problem is another concern, which is that Hawking's analysis has as a premise that explicitly quantum-gravitational effects can be neglected (at least until the decaying hole approaches Planck size).  While this may seem plausible, we now know that it cannot be a foregone conclusion, for there are simple dimensional arguments showing that
quantum-gravitational effects may very well be of a size to seriously alter Hawking's predictions.
These, together with the trans-Planckian problem, mean that we cannot have confidence in any treatment of the quantum effects of black holes without an understanding of at least some aspects of quantum gravity.

In this section, I will review the main elements of Hawking's analysis, with an emphasis on these foundational concerns.  I shall also go over some of the indirect arguments which have been offered in support of Hawking's conclusions:
intuitive pictures about pairs of virtual particles near event horizons;
suggestions that they are necessary to avoid thermodynamic paradoxes; appeals to CPT invariance, etc.  (For further treatment of such matters, see the review article~\citep{Helfer:2003va}.)
It should be clear that no such arguments could really resolve the trans-Planckian problem, or the question of whether 
explicit quantum-gravitational effects should be included.  In fact, some of the arguments are simply incorrect, some would require radical changes in physics, and the others have resisted attempts to make them precise.  In objective terms,
this can be read as evidence against Hawking radiation as easily as evidence for it:  attempts to argue for the radiation seem to uniformly run into difficulties at key steps, and perhaps that is Nature trying to tell us something.
Thus while it is certainly possible that black holes emit thermal radiation, it would be unwise to assume that they necessarily do; there could be some other link between black holes and thermodynamics which results from the physically correct solution of the trans-Planckian problem.

I have gone to some length to outline the inadequacies of the present treatment of quantum theory and black holes,
but I would like to emphasize that Hawking's work does 
strongly suggest that something very deep does link thermodynamics, gravity and quantum theory.  The difficulty is that we know the present treatment is wrong and that we will be unable to be confident in any analysis until we know more about quantum gravity.
While in one sense this is certainly a negative statement, in another it reveals opportunities far deeper than those which would apply had the treatment been adequate.
Yet there is no disguising the fact that this means there is little hope of knowing what work is correct 
and what is not until we have a breakthrough on quantum gravity.

In these circumstances, it is essential that those making arguments about Hawking radiation (and related concepts, such as the ``information paradox'' and  holography) examine which parts of their analyses depend on the problematic elements of the theory.  If, for instance, an analysis only depends on the assertion that black holes radiate, but not on the details of the radiative mechanism, the results evidently have a degree of robustness.  On the other hand, for the deepest progress one wants analyses which do {\em not} have this robustness, for one wants to be able to 
confront the trans-Planckian problem and to
discriminate among the different possible resolutions to it.

\subsection{The Predictions and Their Scale}

Before looking at the analysis in detail, it will be helpful to have a sense of the scale of the effects we are concerned with.

According to Hawking, a spherically symmetric black hole will radiate at the {\em Hawking temperature} $T_{\rm H}=1/8\pi M$ in natural units, where $M$ is its mass.  For stellar-mass or greater holes this is very small:
\begin{equation}
  T_{\rm H}=6.2\cdot 10^{-8} \left(\frac{M_\odot}{M}\right) \, {\rm K}\, ,
\end{equation}
where $M_\odot\simeq 2\cdot 10^{33}\, {\rm g}$ is the mass of the Sun.
Thus, unless we find ``mini'' black holes, there is little prospect for verifying its presence experimentally.

The hole's luminosity will be given by a Stefan--Boltzmann law $L=\sigma T^4_{\rm H}A_{\rm eff}$, where $\sigma$ is the Stefan--Boltzmann constant appropriate to the field species and $A_{\rm eff}$ is the effective radiating area.  It is not a bad approximation to take $A_{\rm eff}=4\pi (2M)^2=16\pi M^2$.  Notice that this means {\em we have a black body whose linear dimension is of the same scale as the wavelength of the dominant quanta at the temperature.  It is therefore in an invariant sense very dim.}  We may see this semiquantitatively by taking 
$T_{\rm H}/L$ to be a characteristic time between the emission of quanta, and $1/T_{\rm H}=8\pi M$ to be the period of a Hawking quantum.  The ratio of these times is
\begin{equation} 
  T_{\rm H}^2/L\simeq 4\pi/\sigma \, .
\end{equation}
Since $\sigma$ is typically a small number, the prediction is that the hole occasionally emits quanta.  It would be more accurate to say it {\em flickers} than that it glows.

The main point of the above computation however is to emphasize how strongly quantum the process is.  The emission of radiation is not simply supposed to be due to quantum effects:  it is quantum in its appearance as well.

\subsection{Hawking's Analysis}

With the background already given, it is not hard to follow the key elements of Hawking's analysis and to appreciate the trans-Planckian problem.  As these points are present even in the simplest case (spherical symmetry and scalar field), we treat those.

So consider a spherically symmetric space--time containing a bounded distribution of matter which is initially dispersed (enough so that we need need not consider strong-field gravitational effects)
and collapses to form a black hole, and a minimally coupled massless scalar field $\phi$ on this background.  We specify the state $|\Psi\rangle$ of the field in terms of its content at early times; it will turn out that the results are largely independent of the particular state, and so one can think of $|\Psi\rangle$ as the state which appears to be vacuum $|0_{\rm p}\rangle$ (as far as the $\phi$ field goes) in the distant past.

We wish to analyze the content of the state as measured by observers far from the incipient hole at late retarded times.  In this regime, the curvature is small and the particle-content of a state is physically well-defined (except for those of extremely low wavelengths, wavelengths of order the distance to the collapsing matter or more).  
We may therefore use the formalism of Bogoliubov transformations to compute the state's appearance in the future.  We saw in section \ref{BTsec} that if the state appeared to be vacuum in the past, in the future it will be a Gaussian amplitude of particles:
\begin{equation}
|\Psi \rangle =\mbox{(normalization)}\exp \left[-(\alpha ^{-1}\beta /2)a^*a^*\right]\, |0_{\rm f}\rangle\, .
\end{equation}
Our task is thus to compute the Bogoliubov coefficients, and especially to focus on the modes for which $\beta$ is non-zero.

To work out the Bogoliubov coefficients, first resolve the field into spherical harmonics:
$\phi =\sum _{l,m}\phi _{l,m} Y_{l,m}$, and then consider the reduced wave equation for each components $\phi _{l,m}$, in the two-dimensional space where the angles have been factored out.  These reduced equations will, as usual, consist of a part independent of $l$ and $m$ together with a centrifugal potential $l(l+1)/r^2$.
Note that the space--time geometry exterior to the matter is characterized by a single dimensionful quantity $M$ (the mass).

Now imagine starting with data for one of these for a wave-packet given near $\scrif$, with nominal angular frequency $\omega$ around a nominal retarded time $u$,
and propagating this backwards in time towards the past.
(This packet is just a c-number field mode.)
As the packet moves inwards, it will be partially dispersed, and partially reflected by the potential.  
In fact, the centrifugal terms (measured at a few Schwarzschild radii) go like $l(l+1)/M^2$, whereas the kinetic terms go like $\omega /M$, and it turns out that most of the wave is reflected for $\omega \lesssim l(l+1)/M$.  This reflected portion of the wave propagates through a static space--time only and so does not give rise to interesting effects.  In fact it is the s-wave, $l=0$, sector, which is almost entirely responsible for Hawking radiation.

Let us continue following the transmitted portion of the wave backwards in time, or more precisely, let us consider a family of such wave packets at increasing nominal retarded times $u$.
Relative to any fixed observer near the horizon, these packets appear to blue-shift by a factor $\sim\exp +u/4M$.  Thus one quickly reaches a point where geometric optics applies.  The propagation backwards through the remaining space--time, including through the region where matter is present, will be governed by geometric optics and the mapping of surfaces of constant phase $\vv (u)$ discussed in section \ref{sphol}.

It is precisely the computation of $\alpha$ and $\beta$ in the geometric-optics approximation, with $\vv (u)\sim -\exp -u/4M$, which is the central part of Hawking's paper.  While one can write the analytic formulas explicitly in terms of the Gamma function, it is not hard to understand their main properties without a detailed computation.  For angular frequencies $\omega \gg 1/4M$ the main effect is to blue-shift the packets to angular frequencies $\omega /\vv '(u)$.  On the other hand, for $\omega \lesssim 1/4M$, the oscillations in the packet change over the same scale as does $\vv '(u)$, and one gets a significant dispersion; in particular, there are contributions to the $\beta$'s as well as the $\alpha$'s.  We can thus see that particle production is predicted to occur for s-wave modes of angular frequencies $\lesssim 1/4M$.
Hawking computes the spectrum in detail, and shows that it is thermal with temperature $T_{\rm H}=1/8\pi M$.  (See also \citet{Wald:1975kc}.)  We shall not need this, but it is also not entirely a surprise, for there is only one dimensionful parameter in the system, the mass.

Thus what we find is that a field mode of finite angular frequency $\omega$ in the future (near $\scrif$) has, in the past (near $\scrip$), an exponentially blue-shifted distribution of frequencies, as measured by the coefficients $\alpha$ and $\beta$.  If $\omega\gtrsim (4M)^{-1}$, then the distribution is fairly sharply peaked, and the main contribution to $\alpha _{\acute\omega \omega}$ is for $\acute\omega\simeq \omega /\vv '(u)$; one has $\beta _{\acute\omega \omega}\simeq 0$.  On the other hand, for $\omega\sim (4M)^{-1}$ one has significant contributions to both $\alpha _{\acute\omega \omega}$ and $\beta _{\acute\omega \omega}$ for $\acute\omega$ within perhaps an order of magnitude of $\omega /\vv '(u)$.

We can now see why the initial state is not very important.  Since if we examine field modes of finite angular frequency $\omega$ in the future, they appear terribly blue-shifted in the past, only detecting the deep ultraviolet content of the state in the past, whose structure is governed by the Hadamard asymptotics, independent of the particular state.

Where do the Hawking quanta come from?  Their precursors, in the distant past, are evidently coded in field modes of 
angular frequencies $\sim (4M)^{-1}/\vv '(u)\sim (4 M)^{-1}\exp +u/4M$.  These modes are unoccupied in the distant past; what is supposed to create the particles is the {\em non-linear} distortion of the retarded time $u$ relative to the advanced time $v$, a distortion which mixes positive- and negative-frequency modes and thus changes the sense of whether modes are occupied or not.
It would be correct to say that the Hawking quanta have their origins in vacuum fluctuations in the distant past.

It is important to appreciate that this particle production is non-local.  We can only confidently speak of identifying particles in regimes in which we have a clear enough sense of time to resolve the field into positive and negative frequencies, and, while these do exist in the future and in the past, there is no clear physically justified means of interpolating between these.

\subsection{The Argument of Bekenstein and Mukhanov}

\citet{Bekenstein:1995ju} gave a simple argument (whose importance was emphasized by \citet{Ashtekar:1998}) which shows that quantum-gravitational effects could easily be of such a magnitude as to completely alter Hawking's predictions.

Consider a spherically symmetric black hole, and suppose that quantum gravity only allows the area to change in multiples of $\alpha l_{\rm Pl}^2$, where $\alpha$ is a constant.  While there is no strong argument that this would be the case, it is the sort of effect which might very plausibly arise, and it is an attractive hypothesis if we wish to think of the black hole's area as an information-theoretic entropy.

We have $A=4\pi (2M)^2=16\pi M^2$, and so $\Delta A=32\pi M\Delta M$, or $\Delta M=\Delta A/32\pi M=(n\alpha /4)T_{\rm H}$.  In other words, the hole's mass-energy can only change by discrete units, and so Hawking quanta can only be emitted if their energies are one of these units.  If $\alpha\ll 1$, then the allowed lines are very dense, in effect quasi-continuous, and there is little change to Hawking's analysis.  But if $\alpha \sim 1$, only a fraction of the Hawking quanta will have the correct energy, and the radiation would be significantly curtailed; for $\alpha\gtrsim 10$ there would be almost no Hawking radiation.

Thus quantum-gravitational effects could, although they need not, completely alter Hawking's analysis, and we cannot have confidence in the prediction of black-hole radiation without understanding some features of quantum gravity.\footnote{In loop quantum gravity models, it is shown that while area is quantized the spacing is very fine except for Planck-scale holes.}

\subsection{Some Arguments on Hawking Radiation}

It should be clear that no argument within conventional quantum field theory in curved space--time could remove the trans-Planckian problem, or eliminate the possibility that quantum-gravitational effects might seriously alter the prediction of black-hole radiation.  These issues are recognized in {\em non-standard propagation models}, which explicitly aim to save the predictions by altering the propagation rules for the quantum fields (see~\citep{Helfer:2003va} for references and discussion).

One can also try to find indirect arguments in favor of thermal black-hole radiation.  These all run into difficulties at essential points, and it is not clear whether those difficulties are signals that the aim is incorrect or just that the analysis is not deep enough.
I have elsewhere~\citep{Helfer:2003va} critiqued a number of these (claims that the radiation is necessary for the thermodynamic consistency of general relativity and that it follows from Unruh radiation via the equivalence principle), and I shall not repeat these here.  I will briefly comment on some other arguments which have appeared.

{\em Hawking radiation is implied by CPT invariance.}  
There are several versions of this statement in the literature.  As far as I know, they break down into:  (a) interesting but radically speculative arguments to the effect that Hawking radiation and CPT invariance {\em together} form a picture the author finds attractive, but which cannot objectively be considered independent evidence for Hawking radiation; (b) incorrect arguments; (c)~assertions without supporting arguments.  

Examples of arguments of class (a) are those of \citet{Hawking:1976de} and \citet{Hooft:1996tq}.  
While it would be out of place to describe these works in detail here, as indices of their speculative character I'll note that Hawking argues that black holes and white holes ought --- quantum-theoretically --- to be the same thing!  (Since the classical causal structures of black and white holes are clearly distinct, this view implies gross modifications to causal structure via quantum effects, and thus really amounts to speculations about quantum-gravitational effects severely altering classical relativity.)  And 't~Hooft's argument {\em starts} from the premise that a black hole should have the same properties as a special-relativistic particle.

Arguments of type (b) assert that because black holes form due to processes $\mbox{(particles)}\to\mbox{(black hole)}$ there should be corresponding decay processes $\mbox{(black hole)}\to\mbox{(particles)}$.  This is seriously incorrect, at two levels. 
First, it is hardly a foregone conclusion that the CPT theorem will hold --- or even be formulatable --- in a unified theory of quantum fields and gravity, and indeed experimental searches for CPT violation are of considerable interest as evidence for quantum-gravitational effects. 
And even if the CPT Theorem holds in this context, the (b)-type arguments do not apply it correctly.
The theorem relates processes to their time-reversed (and CP-transformed) versions, so it would relate the black-hole formation process to a {\em white}-hole decay process $\mbox{(white hole)}\to\mbox{(particles)}$.

A correct application of the sorts of ideas underlying this actually provides a pair of alternatives, neither of them really arguments in favor of Hawking radiation (and one tending against it).
If Hawking's notion that black holes radiate completely away is correct, that is,
the process $\mbox{(black hole)}\to\mbox{(particles)}$ exists, and CPT applies, then the process $\mbox{(particles)}\to\mbox{(white hole)}$ must be possible, that is, a white hold could
be formed from a non-problematic initial state.
This would be almost universally considered highly counterintuitive, and so tends to make Hawking radiation less plausible. 

If, on the other hand, Hawking's notion that black holes decay is correct but they leave remnants,
and CPT applies,
the forward process would be  $\mbox{(black hole)}\to\mbox{(particles + remnant)}$, and so the reverse would be $\mbox{(particles + seed)}\to\mbox{(white hole)}$, where ``seed'' is the CPT-reversal of a remnant.  This would say that white-hole formation via this mechanism would require a seed, and, if such objects are sufficiently exotic then the white-hole formation mechanism itself might be plausible.

{\em Analytic-continuation arguments.}  These include appeals to the very beautiful structure of the analytically-continued Schwarzschild solution which arises when it is used to model a thermal state at the Hawking temperature, and Hawking's Euclidean quantum gravity program.

First and foremost, these arguments do not address any of the difficulties associated with Hawking radiation.  Instead, what they do is provide elegant formal mathematical structures which however it is hard to link in detail with the physical processes.  (Even leaving aside the serious doubts about whether physically relevant solutions admit the requisite analytic continuations, in general analytic continuation is a highly nonlocal procedure.)

Hawking \citep{Hawking:1996jh} has argued that Euclidean quantum gravity is simply not meant to be anything more than a calculational tool for working out transition probabilities, and thus that it is wrong to fault its internal mathematics for not connecting clearly to the physical world.  If one holds this position, however, one cannot argue that at this level Euclidean quantum gravity provides an answer to the trans-Planckian problem, since what is required to resolve that is precisely a detailed explication of what happens to the field modes. 
More generally, while a view like Hawking's has some pragmatic justification
in particle scattering theory, where it is very hard to probe the dynamics of the processes and we by and large settle for the overall transition probabilities, in gravitational physics we are very definitely interested in the details of the dynamics.\footnote{Even in the special-relativistic case, one would like to treat the full dynamical structure of the quantum theory.  The restriction to scattering problems has been enormously fruitful, but one would certainly like to go further.}  It is true that there may well be quantum-gravitational corrections which limit the utility of the classical space--time concept in describing these, but if so we want to be able to say as explicitly as possible what these are.  

In addition, of course, any quantum gravity program is speculative, and none has really been developed satisfactorily.

{\em Hawking radiation arises from particles tunneling through the horizon.}
Essentially this same language has been used to summarize several different physical arguments.

When it is applied to describe Hawking's original argument, it is seriously misleading:
(a) Most grossly, it suggests that there is an {\em acausal} element to the mechanism, with some sort or physical process
propagating outwards across the horizon.  Nothing could be
further from the truth.
(b) Hawking's prediction is that black-hole radiation is a {\em quantum-field-theoretic phenomenon in which the particle number changes}, from initially vacuum to a steady flux.  This is clearly not the same as a quantum-{\em mechanical} tunneling process conserving particle number.
(c) One can try to repair these defects by asserting that it is ``virtual'' particles which propagate across the horizon and tunnel.  The difficulty with this is that the term ``virtual particle'' is in this context so vague that it conveys little information.

\citet{Parikh:1999mf} give a very interesting attempt to make explicit a tunneling model.  However, just how this connects with quantum field theory in curved space-time is unclear.  Work on this would be worthwhile, especially because the Parikh--Wilczek ideas treat the back-reaction as an essential element of the physics, in contrast to the usual Hawking picture.

\section{Black Holes, Quantum Fields, \protect\\ Measurements and Interactions}\label{moreqft}

I have shown above that the usual theory of quantum fields in curved space--time is not adequate for treating black holes.  This is surprising, as the theory is both natural and attractive.
We should take this inadequacy very seriously, as it is giving us valuable information about what issues must be confronted in reconciling quantum theory with gravity.  And since that is a deep problem, it is worthwhile exploring the information we have about it --- the nature of the failures of the usual theory --- in some detail.  

I shall first give a simple and suggestive, but not conclusive, argument to the effect that a black hole, or more precisely what appears to be an incipient black hole, has a sort of boundary beyond which it becomes an essentially quantum object. 

I shall then give three analyses which bear on the trans-Planckian problem and, more generally, on the theory's treatment of ultraviolet virtual effects.  In each of these virtual effects are promoted to real ones.  (From one point of view, this is hardly surprising; a great many of the interesting effects in special-relativistic quantum field theory come from real consequences of virtual effects.)  The first case involves a computation of the results of sequences of measurements of quanta in the Hawking effect; the second (closely related to the first) takes into account quantum-field-theoretic nonlinearities in the Hawking effect; and the third is an estimate of a Casimir effect of a black hole.
In each of these cases, the result is unacceptable, and indeed absurd.  We shall find that in each case the theory predicts {\em real} trans-Planckian effects.

While these failures are particularly blatant in the case of black holes, on account of the exponentially increasing red-shifts affecting mode propagation, there is no reason to think they are special to black holes.  Even without these exponential increases, sufficiently high red-shifts would give rise to problems in all of these cases.  Indeed, the Casimir result is problematic for even modest red-shifts.
I would therefore suggest that there is a real possibility that the theory of quantum fields in curved space--time, as natural as it does appear, may turn out to be in need of serious modification.  

\subsection{Measuring Geometry Near a Black Hole}

Let us imagine a spherically symmetric distribution of matter which is collapsing and will ultimately form a black hole.  Let us ask how well distant observers may measure the geometry of the space--time in the vicinity of the incipient horizon; we shall consider only spherically symmetric measurements here.

In particular, we have seen that a key element of this geometry is the fractional rate of red-shift $v''/v'$.  This could, by the universality described above, be measured by looking at signals emitted by any objects falling into the hole. Suppose such signals are wave-packets emitted with known (nominal) frequency in the object's local frame (they could be particular spectral lines), and they are received at $\scrif$ with (nominal) angular frequency $\omega$.  Thus measurements of the rate of change of $\omega$ can be used to infer $v''/v'$.  

Now we must have $\omega\gtrsim |v''/v'|$ in order to resolve the geometry in question. If the system is approaching a Schwarzschild solution of mass $M$, this means we should have $\omega\gtrsim c^3/GM$.
On the other hand, the angular frequency in the emitted object's frame must be $\sim k\omega /v'$, where $k$ depends on the velocity of the emitting object as it moves to cross the horizon, but $1/v'$ increases exponentially.  We presumably must have $k\hbar \omega /c^2v'\leq M_{\rm Pl}$, the Planck mass. 
Combining the inequalities, we find $k\hbar c/GMv'\lesssim M_{\rm Pl}$, which is
\begin{equation}
  v'\gtrsim kM_{\rm Pl}/M\, .
\end{equation}
This inequality represents a limitation on the portion of the geometry which can be measured by distant observers.
Formally, it defines the part of the tangent bundle which is accessible to measurement.  Because $v'$ approaches zero so rapidly, however, it means that for all practical purposes the close neighborhood of the event horizon cannot be investigated; we cannot operationally treat it as having a classical geometry.

\subsection{Measurements of Hawking Quanta and Their Precursors}

While a great deal of work has been done on quantum field theory in curved space--time, relatively little attention has been given to the investigation of the effects of quantum measurements, that is, the interesting possibilities arising from measuring different, in general non-commuting, operators.  Yet this is a basic element of quantum theory, and any attempt to assess the plausibility of an analysis in curved space--time should consider the predictions of such measurements.

I am going to describe here the results of measurements of Hawking quanta and their precursors, according to conventional quantum theory~\citep{Helfer:2004jx}.  What we shall find is that, because the measurement process introduces a coupling between the measuring device and the field modes, measurements can promote the trans-Planckian problem from a virtual to a real one, where it is manifestly unacceptable.

{\em Measurements and Unruh Detectors.}
Before doing this in the Hawking case, however, I want to describe something similar in the Unruh process.  Let $\gamma (s)$ be the world-line of a detector responding to a linear field in Minkowski space.  If the detector is strictly localized to the world-line, then it cannot really be said to be a particle detector, but the question of just what name is best for the quanta the detector responds to will not be important here, neither will strict localization to the world-line.  We will also assume that the detector was initially inertial, but then is smoothly accelerated.

The accelerating detector measures what appears, along  the accelerating world-line, to be a number operator $n=n(\gamma ,\omega , s_0,s_1)$ corresponding to measurements of field modes of a angular frequency $\omega$ over a particular sampling time $s_0\leq s\leq s_1$; it is determined by 
resolving the field $\phi (\gamma (s))$ along the world-line into Fourier components (with respect to the proper time $s$), and following the usual constructions as if $s$ were ordinary time.  

Now let us suppose the state is initially the vacuum $|0\rangle$, and then a measurement of $n$ is made.  Then the state will be projected into an eigenstate of $n$.  This will certainly not be the vacuum, and in fact it will be a superposition of excited states.  Typically, these will contain quanta of energies $\sim \omega t_a\dot\gamma ^a(s)$ with respect to the frame with timelike vector $t^a$, where $s_0\leq s\leq s_1$.  (This will occur even if the measurement of $n$ yields zero.)

Note that if the detector has been accelerated to a very high boost relative to $t^a$, then detection will result in the state containing superpositions of very high energies.  It is natural to ask where these energies come from.  In some sense, they must come out of the measurement process, and the high relative energy of the detector with respect to the frame described by $t^a$.

{\em Measurements and Hawking Detectors.}
Imagine two observers who follow trajectories each everywhere far from an object collapsing towards a black hole.  Each carries a detector, which for convenience one may think of as a camera containing a photographic plate, and the shutter of the camera will be controlled to open at certain times.  What will be critical in the subsequent analysis will be to distinguish between simply exposing the plate (which is to say, allowing a coupling to certain field modes) and inspecting it (actually doing a quantum measurement).  Let the initial state of the field be $|\Psi\rangle$, say the vacuum in the distant past.

The plates of the first observer, call her H, are sensitive to quanta around the characteristic Hawking angular frequency $T_{\rm H}$.  Her camera is set to have its shutter open during an interval of retarded time $\Delta u$ around a time $u$ late enough so that Hawking quanta are expected there.  
The plates of the second observer, call him UE, are sensitive to quanta at the angular frequency of the precursors to those measured by H, that is, at $(v'(u))^{-1}T_{\rm H}$  --- UE is for ``ultra-energetic.''  His camera is set to have its shutter open in an interval around the advanced time $v(u)$ at which those precursors are moving inwards towards the origin.

In fact, the cameras simply enable H and UE to measure the number operators of quanta, H measuring Hawking quanta and UE the quanta at the frequency of the precursors.
We now consider what happens when H and UE make their measurements, that is, examine the plates.  The results depend critically on which order this is done in.

\begin{wrapfigure}{r}{0.4\textwidth}
\vspace{-1.2em}
  \begin{flushright}
{\includegraphics[width=.35\textwidth]{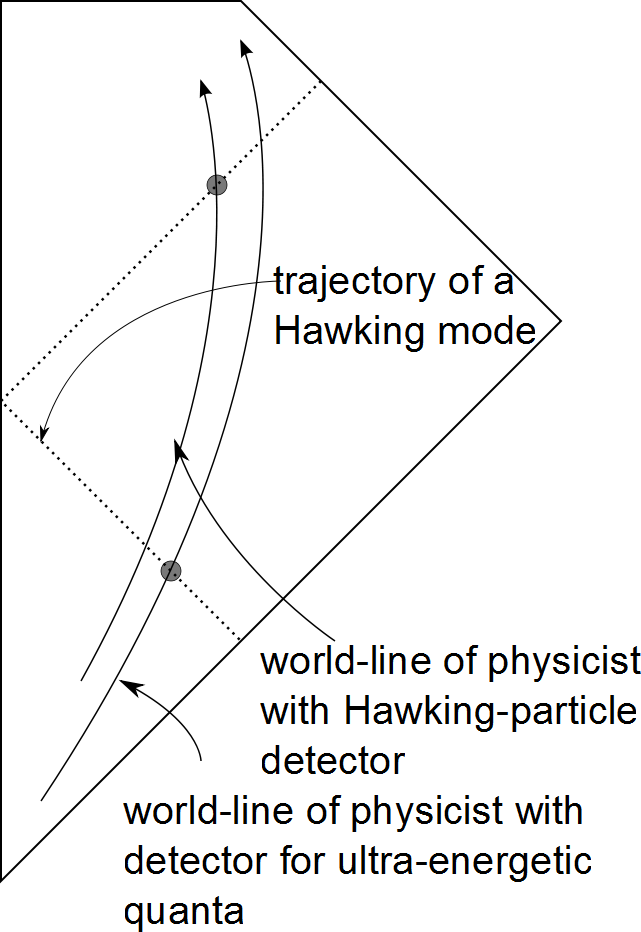}}
\makebox[.35\textwidth]{$\ $ }
\vspace{-.1em}
\parbox{.35\textwidth}{4.  If observer H measures Hawking modes first, observer UE has a finite probability of detecting an ultra-energetic excitation.  The dots indicate where the shutters are open.  The relative distances of the detectors from the incipient hole are unimportant.}
\end{flushright}
\label{fig:HUE}
\end{wrapfigure}

If UE goes first, nothing remarkable happens.  UE measures the particle-number for an ultra-energetic mode in the past, but this mode was not excited in the state $|\Psi\rangle$.  In other words, the state is already an eigenstate of that operators.  Thus UE's measurement returns the value zero.  Then H may or may not record a number of Hawking quanta, with probabilities given according to the usual Bogoliubov transformation.

But if H goes first, things are very different.  She will record a finite number of Hawking quanta.  The precise number does not matter in this analysis, and the most likely number (if $\Delta u$ is not too long) is zero, so let us say that is what she gets.  Then the state is projected to an eigenstate of her number operator.

It is possible to work this out exactly, but a rough computation will convey the main idea.  Let us ignore dispersive effects in the propagation of the mode in question, so the annihilation operator $a$ for her mode can  be written as $a=\alpha b+\beta b^*$
in terms of the annihilation and creation operators for UE's mode. 
Because H's mode is a Hawking one, the Bogoliubov coefficients $\alpha$ and $\beta$ have comparable magnitudes.
Then the state $|0_{\rm H}\rangle$ which H's measurement has projected $|\Psi\rangle$ to is characterized by $a|0_{\rm H}\rangle =0$, whence
\begin{equation}
  |0_{\rm H}\rangle =\mbox{(normalization)}\cdot
  \exp [ -(\beta /2\alpha ) b^*b^*] |0_{\rm UE}\rangle
\end{equation}
in terms of the eigenstates of UE's number operator.  That is, the result of H's measurement, the state $|0_{\rm H}\rangle$, appears in terms of the number basis in the past to be a superposition of excitations of ultra-energetic modes.
If now UE makes a measurement, he will with finite probability record a number of these ultra-energetic quanta.

This result is unacceptable and indeed absurd, because the energies $(v'(u))^{-1}T_{\rm H}$ of the quanta supposedly measured by UE grow exponentially quickly, passing not only the Planck scale but indeed the mass--energy of the collapsing object (and even of the observed Universe).

\subsection{Interacting Quantum Fields}

Virtually all of the specific computations of quantum fields in curved space--time have been done in the case of linear field theories, because we do not yet have a practical understanding of how to treat the nonlinear case.  The concern has therefore been raised that Hawking's analysis might be inadequate because of its neglect of nonlinear effects.\footnote{In attempts to model astrophysical black holes there have also been what one might call phenomenological models, where certain species are assumed to be produced by the (linear) Hawking mechanism, and then interact as they propagate outwards (e.g.,  \citep{MacGibbon:1990zk}).  However, this does not address the concern that nonlinearities have not been taken into account at a fundamental level.}

The most concrete attempt to respond to this was a very interesting paper by \citet{Gibbons:1976es}.  They sketched an argument to show that at late time all the Feynman propagators of the theory would have  a periodicity in imaginary time resulting in the field theory being at the Hawking temperature.  However, the argument was not complete.  For one thing, the authors did not really spell out the issues involves with ``dressing'' the in and out states (and we shall see shortly this is essential).  For another, while they emphasized the importance of showing that an initially vacuum state would dynamically equilibrate to a thermal one, it was in their analysis unrealistically hard to track this behavior, and instead they worked with the ``Hartle--Hawking'' state, which in effect builds in the assumption that an equilibrium has been reached and one can neglect all the early-time physics.  In particular, issues like the trans-Planckian problem cannot be properly investigated with this assumption.

One also sometimes hears the suggestion that perhaps {\em asymptotic freedom} (the running of couplings towards zero at higher energy scales) may solve this problem.  As far as I know, however, no careful argument has been offered to this effect, and it is far from clear how one could be made to work.

I am going to outline a computation of a nonlinear quantum-electrodynamic correction to Hawking's analysis~\citep{Helfer:2005wy,Helfer:2005wz}.  It will be an effect which is 
{\em first-order in the electric charge.}  (In special-relativistic computations in scattering theory, one only gets contributions in even powers of the charge, that is, integral powers of the fine-structure constant; however, here we will get an amplitude which is first-order.)  I am in fact going to choose the particular amplitude so that we do not have to confront renormalization theory beyond normal ordering.  All of the interesting effects will be due to the change in ``dressing'' in the future compared to the past.

Let us begin by considering the situation in the distant past, when the matter is dispersed, and the space--time is to good approximation Minkowskian.  Then the quantum-electrodynamic Hamiltonian is
\begin{equation}
H=H_{\rm b}+H_{\rm int}
\end{equation}
where $H_{\rm b}$ is the ``bare'' Hamiltonian, describing the linear fields, and
\begin{equation}
  H_{\rm int}=-e\int _\Sigma \tilde\psi \gamma ^a\Phi _a\psi
\end{equation} 
is the interaction.  Here $\psi$ is the charged field operator and $\Phi _a$ is the electromagnetic potential operator, and $\Sigma$ is the Cauchy surface of interest.

The first thing we must consider is that the vacuum is altered by the interaction, the dressed vacuum being given, to first order in the electric charge, by
\begin{equation}
 |0_{\rm d}\rangle =(1-H_{\rm b}^{-1}H_{\rm int})|0_{\rm b}\rangle\, ,
\end{equation}
according to standard perturbation theory.  In this expression, the factor $H_{\rm int}|0_{\rm b}\rangle$ will contain bare electron, positron and photon creation operators acting on $|0_{\rm b}\rangle$; one may say that the dressed vacuum contains ``bubbles'' of bare electrons, positrons and photons.\footnote{
However, it should be emphasized that $|0_{\rm d}\rangle$ really is the physical vacuum, no-particle, state, and these bubbles are an artifact of examining the state in terms of the bare particles, which are mathematically convenient but not of direct physical relevance.}
These bubbles involve modes with arbitrarily large wave-numbers (which ultimately contribute to the ultraviolet divergences of the theory).  However, a key point is that the spatial wave-numbers of each vacuum triple sum to zero.  
This is represented diagrammatically by the particle lines terminating at common vertices.

\begin{wrapfigure}{r}{0.35\textwidth}
  \begin{flushright}
{\includegraphics[width=.3\textwidth]{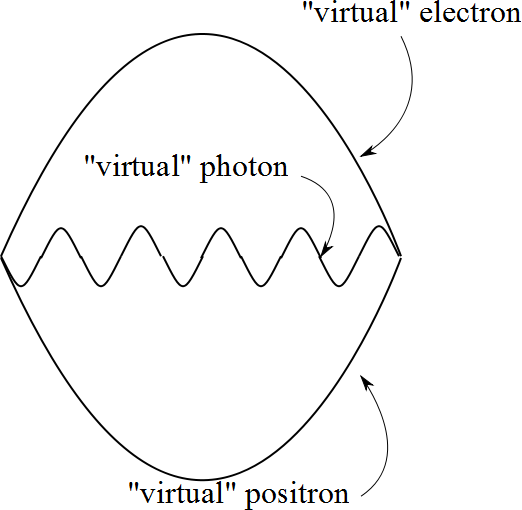}}
\makebox[.3\textwidth]{$\ $ }
\parbox{.3\textwidth}{5. The quantum-electrodynamic interaction modifies the vacuum by creating ``bubbles'' of virtual bare quanta.}
\end{flushright}
\label{fig:vacbubble}
\end{wrapfigure}

The next issue is how to extract the physically significant information about the state in the future.  In special-relativistic theory, one would generally do this by analyzing its particle content.  This involves technical issues which would take us too far afield.  Instead we shall concentrate on the field aspect, and only give heuristic comments about the particle content.

A basic element of quantum electrodynamics is
the {\em interaction vertex}, which is $\langle 0_{\rm d}|{\tilde\psi}(k')\Phi (k)\psi (k'')|0_{\rm d}\rangle$, where $k'$, $k$ and $k''$ are mode labels which include both momentum and polarization information.  We will compute this for certain modes.  The point will be that any difference we find between the black-hole and the Minkowskian cases will represent an excitation of the black-hole case relative to the Minkowskian vacuum --- taking the difference amounts to renormalizing.

The interaction vertex for the particular modes in question will be 
\begin{equation}\label{vert}
\langle 0_{\rm d}|{\tilde\psi}_{\rm f}^+(k')\Phi _{\rm f}^- (k)\psi _{\rm f}^+ (k'')|0_{\rm d}\rangle\, ,
\end{equation} 
where the postscripts indicate positive or negative frequencies in the future.  For the modes, we take $k$ to correspond to a characteristic Hawking photon in the future, and $k'$, $k''$ will describe field modes propagating everywhere far from the incipient hole.  (A more precise specification of 
$k'$ and $k''$ will be given in a moment.)

First-order contributions to the vertex function~(\ref{vert}) can potentially come from three places:  perturbations of the bra relative to the bare vacuum, perturbations of the operators relative to the bare ones, and perturbations of the ket relative to the bare vacuum.  Of these, the only non-trivial contribution arises from the perturbation of the ket, because in each of the other cases one has at least one of the bare charged field destruction operators acting on the bare vacuum.  Thus to first order we have
\begin{equation}\label{verte}
\langle 0_{\rm d}|{\tilde\psi}_{\rm f}^+(k')\Phi _{\rm f}^- (k)\psi _{\rm f}^+ (k'')|0_{\rm d}\rangle=
-\langle 0_{\rm b}|{\tilde\psi}_{\rm f}^+(k')\Phi _{\rm f}^- (k)\psi _{\rm f}^+ (k'')H_{\rm b}^{-1}H_{\rm int}|0_{\rm b}\rangle\, ,
\end{equation}
where the operators may be taken to be bare.  Here the factor $H_{\rm int}$ creates vacuum bubbles of bare particles, and $H_{\rm b}^{-1}$ only changes their weightings.  Since $k'$ and $k''$ are chosen to represent modes everywhere far from the collapsing region, the charged-field operators annihilate the bare quanta in the vacuum bubbles.  
Were we in Minkowski space, the field $\Phi ^-_{\rm f}(k)$ would annihilate the bra, and the expression~(\ref{vert}) would vanish.  Here, however, that field, which represents creation of a bare Hawking photon in the future, will in the past represent a precursor, which will have a mixture of positive and negative frequency terms, but will be highly blue-shifted.  Thus $k$, in the distant past, will correspond to an ultra-high frequency wave-vector directed inwards.  

\begin{wrapfigure}{r}{0.4\textwidth}
  \begin{flushright}
{\includegraphics[width=.35\textwidth]{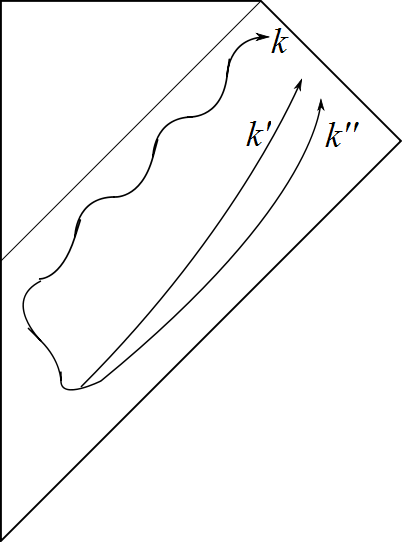}}
\makebox[.35\textwidth]{$\ $ }
\parbox{.35\textwidth}{6.  When the quanta from a vacuum bubble can exchange energy--momentum with the dynamical space--time, resulting in the bubble failing to close and the promotion of the quanta from virtual to real.}
\end{flushright}
\vspace{-2em}
\label{fig:brokenbubble}
\end{wrapfigure}

Now recall that the sum of the spatial wave-numbers of the vacuum triples is zero, so we will get non-trivial contributions to the interaction vertex when $k'$, $k''$ correspond to wave-vectors whose total spatial component is ultra-high and directed {\em outwards}.  We thus see that there will be, in the future, non-trivial contributions to the interaction vertex corresponding to ultra-high frequencies in the charged fields.  

I pointed out above that the analysis in terms of physical particles involves further technicalities, but if we ignore the niceties, then the bra $\langle 0_{\rm d}|{\tilde\psi}_{\rm f}^+(k')\Phi _{\rm f}^- (k)\psi _{\rm f}^+ (k'')$ represents a state 
which is a modification of the in-vacuum by deleting (in terms of the out-regime)
one Hawking photon and creating an ultra-energetic electron-positron pair; the vertex function~(\ref{vert}) can thus be thought of as the amplitude to find this state given that the initial state was the vacuum.  In other words, speaking of particles in this heuristic sense, there would be a finite probability of an ultra-energetic electron-positron pair appearing at late time.  While this particulate interpretation is problematic, the interpretation at the level of three-point functions shows unequivocally that  in the future the state $|0_{\rm d}\rangle$ possesses ultra-high wave-number excitations relative to the Minkowski vacuum.

This result is unphysical and absurd because the characteristic wave-numbers for the charged pair increase exponentially quickly, with $(v'(u))^{-1}$.  
I would like to emphasize that this computation follows very closely the original point of view from which Hawking's predictions were derived, and is conventional as far as both quantum field theory and relativity go.  

\subsection{A Casimir Calculation}

One would like to understand how quantum fields can act as sources for gravity.  In the case of black holes and Hawking-type proposals, this is {\em the back-reaction problem:}  how do the effects of the quantum field change the geometry of the space--time?  If in fact black holes radiate, and if in any usual sense energy is conserved, presumably the black hole (or more accurately, the incipient hole) must lose the energy emitted.  But this must mean that in some sense there is a negative energy flux inwards towards the incipient hole.

We do not have a good understanding of how quantum fields act as sources for gravity.
The most direct analog of a classical source term is the {\em stress--energy operator}.  This is not at all trivial to construct, because the formal expression for it involves products of the field operators at the same point, and these are only {\em distribution}-valued operators and the product of two distributions is not generally defined.  One has to renormalize this formal expression via an infinite subtraction to obtain a well-defined operator.  This, it turns out is always possible for quantum fields in curved space--time, and the result is the {\em renormalized stress--energy operator} $T^{\rm ren}_{ab}$.
On the other hand, there is an ambiguity in the process, because there are no general criteria known for fixing the zero-point of the subtraction, and thus the renormalized stress--energy operator is only determined up to the addition of a conserved c-number term~\citep{Wald:1995yp}.

It is probably fair to say that most theorists regard this as a fairly good state of affairs, and that hopefully at a later date we will have a better understanding of the c-number ambiguity.  I think that more caution is in order.  Even if we defer the issue of the c-number ambiguity, the stress--energy operator is rather problematic, giving generically measures of energy which are unbounded below due to ultraviolet problems~\citep{Helfer:1996my}.  And here I will show that trying to reconcile the c-number contributions with the standard understanding of Casimir effects leads to another serious problem.

If it is possible to argue for a particular system that we have a good way of fixing the 
c-number contributions to the stress--energy, and if the system possesses a lowest-energy state, then we refer the expectation of the stress--energy in that state as a {\em Casimir term}.  Similarly, even if it may not be clear how to fix the c-number contribution to the entire stress--energy, but we can fix the contribution to the Hamiltonian, the lowest energy is the {\em Casimir energy}.  Generally, the most secure computations of Casimir effects occur when we have a mode-by-mode comparison with a known system.

The first and most famous example is due to Casimir, following an exchange of his with Bohr.  Consider the region between two parallel  perfect plane conductors separated by a distance $l$.
In this case, one has a mode-by-mode comparison of the finite-$l$ case with the Minkowski vacuum $l\to\infty$.  It turns out that the Casimir stress--energy is
\begin{equation}\label{Cas}
  \langle 0 |T_{ab}^{\rm ren}|0\rangle =-\frac{\pi ^2\hbar c}{1440 l^4}\left[
  \begin{array}{cccc} 2&&&\\ &1&&\\ &&1&\\&&&0\end{array}\right]
\end{equation}
in a Cartesian coordinate system adapted to the symmetry of the problem.  Of course, the most striking thing about this is that the energy density is negative; Casimir energies may have either sign.
(For questions about the detectability of negative energy densities, see~\citep{Helfer:1998}.)  
That the Casimir energy density comes out to be constant between the plates is a feature of the idealized nature of the problem.  Also its modest size is the result of very delicate calculations:  the renormalized squares of the electric and magnetic fields separately actually {\em diverge} with opposite signs as one approaches the plates.  (The divergence is again due to the idealized, perfect-conductor, boundary conditions; more realistic situations can have large finite energy densities.)
Finally, there is little chance at present for experimentally detecting Casimir energies (they are too small), but there is some prospect for observing Casimir pressures (see e.g. \citep{Lamoreaux:2010,KM:2010}).

Now consider as usual a massless scalar field propagating in a spherically symmetric space--time representing a system collapsing to a black hole.  We shall only be concerned with the ultraviolet asymptotics of this theory, that is, the propagation of very high frequency modes.  (Whether the state is excited in these modes will not matter; it will be most natural to assume these modes are unexcited.)  For high enough frequencies, these modes propagate to good approximation by geometric optics.  

Let us now consider just the s-wave modes which are supported at late retarded times --- these are like the Hawking modes, but we are here interested in the ultraviolet regime, where there are few Hawking quanta.  Since $T_{\rm H}$ is the only relevant characteristic scale, we may say we are interested in modes here of angular frequencies $\omega\gtrsim T_{\rm H}$.  Each of these modes will have a vacuum stress--energy contribution.

We may compute the s-wave contribution $\Phi _{\rm s}$
to the 
vacuum flux $\int T_{ab}^{\rm vac} n^an^b d^2S$ over a large sphere.  The s-wave fields near $\scrif$ has the form $\phi _0(u)/r$, where $\phi _0(u)$ is effectively a quantum field in one dimension~\citep{Helfer:2003va}.  Using a standard result from this theory, we find
\begin{eqnarray}
  \Phi _{\rm s}&=&\langle 0|\int n\cdot\nabla\phi \Bigr|_{\rm s} 
n\cdot\nabla\phi \Bigr|_{\rm s} d^2S|0\rangle\nonumber\\
  &=&4\pi\langle 0|\partial _u\phi _0\partial _u\phi _0|0\rangle
  \nonumber\\
&=&(2\pi )^{-1}\int _{T_{\rm H}}^\Lambda\omega d\omega\nonumber\\
&=&(4\pi )^{-1}(\Lambda ^2-T_{\rm H}^2)\, ,
\end{eqnarray}
where $\Lambda$ is an ultraviolet cut-off.
That is, the vacuum energy due to s-wave modes passing through an interval of retarded time of length $du$ will be $(4\pi )^{-1}(\Lambda ^2-T_{\rm H}^2) du$.

On the other hand, those same modes originated at $\scrip$ with their frequencies blue-shifted by a factor $(v'(u))^{-1}$, so the energy inwards at $\scrip$ will be $(4\pi )^{-1}(v'(u))^{-2}(\Lambda ^2 -T_{\rm H}^2) du$.  Thus if we mode-by-mode compare the vacuum energies, we find
\begin{equation}\label{Casdif}
  E_{\rm future}-E_{\rm past} =(4\pi )^{-1}(1-(v'(u))^{-2})(\Lambda ^2-T_{\rm H}^2) du\, ,
\end{equation}
which is grossly negatively divergent.  In other words, {\em the usual arguments about Casimir energies lead to the conclusion that the black hole absorbs a divergent Casimir contribution.}

This conclusion is again unacceptable and absurd.  
The problem is, however, to put one's finger on what is wrong.  Simply rejecting this argument wholesale would seem to require also rejecting more familiar instances of Casimir calculations.  And while I have indicated above that there are some questions about those, it is hard to believe they are entirely wrong.

In fact, the divergence we have found here
does not depend at all on having an incipient black hole; it is simply a consequence of the modes suffering a red-shift in their passage through the space--time.  (That, in eq.~(\ref{Casdif}), the factor $v'(u)$ tends rapidly to zero is not needed for there to be a divergence; one only needs $v'(u)\not=1$.)  So this problem with Casimir stress--energy is really a generic problem for quantum fields in curved space--time.

\section*{Conclusion}

The theory of black holes has given us results of extraordinary depth and beauty.  It is the resulting pushes to
treat still more difficult and fundamental issues which have led to the problematic points discussed here.  These points show that black holes are linked to some of the deepest problems in physics, and we have yet much to learn from them --- if we are up to the challenge.

\begin{theacknowledgments}
It is a pleasure to thank the organizers and participants of BSCG XIV for a stimulating and enjoyable time.
\end{theacknowledgments}

\bibliographystyle{aipproc}


\begin{thebibliography}{57}
\expandafter\ifx\csname natexlab\endcsname\relax\def\natexlab#1{#1}\fi
\providecommand{\enquote}[1]{``#1''}
\expandafter\ifx\csname url\endcsname\relax
  \def\url#1{\texttt{#1}}\fi
\expandafter\ifx\csname urlprefix\endcsname\relax\def\urlprefix{URL }\fi
\providecommand{\eprint}[2][]{\url{#2}}

\bibitem[Doeleman et~al.(2009)]{Doeleman:2009te}
S.~Doeleman, et~al., {Imaging an Event Horizon: submm-VLBI of a Super Massive
  Black Hole} (2009), \eprint{arxiv:0906.3899}.

\bibitem[Penrose and Rindler(1986)]{Penrose:1986ca}
R.~Penrose, and W.~Rindler, \emph{{Spinors and space--time. Vol. 2: Spinor and
  twistor methods in space--time geometry}}, Cambridge University Press, 1986.

\bibitem[Carter(1979)]{Carter:1979}
B.~Carter, \enquote{The general theory of the mechanical, electromagnetic and
  thermodynamic properties of black holes,} in \emph{General relativity: an
  Einstein centenary survey}, edited by S.~W. Hawking, and W.~Israel, Cambridge
  University Press, 1979, pp. 294--369.

\bibitem[Booth(2005)]{Booth:2005qc}
I.~Booth, \emph{Can. J. Phys.} \textbf{83}, 1073--1099 (2005),
  \eprint{gr-qc/0508107}.

\bibitem[Penrose(1972)]{Penrose:1972}
R.~Penrose, \emph{Techniques of differential topology in relativity}, vol.~7 of
  \emph{C.B.M.S. Regional Conf. Ser. in Appl. Math.}, S.I.A.M., 1972.

\bibitem[Israel(1989)]{Israel:1987ae}
W.~Israel, \enquote{Dark Stars: The Evolution of an Idea,} in \emph{Three
  hundred years of gravitation}, edited by S.~W. Hawking, and W.~Israel,
  Cambridge University Press, 1989, pp. 199--276.

\bibitem[Oppenheimer and Snyder(1939)]{Oppenheimer:1939ue}
J.~R. Oppenheimer, and H.~Snyder, \emph{Phys. Rev.} \textbf{56}, 455--459
  (1939).

\bibitem[Hawking and Ellis(1973)]{Hawking:1973uf}
S.~W. Hawking, and G.~F.~R. Ellis, \emph{{The large scale structure of
  space-time}}, Cambridge University Press, 1973.

\bibitem[Ashtekar and Krishnan(2004)]{Ashtekar:2004cn}
A.~Ashtekar, and B.~Krishnan, \emph{Living Rev. Rel.} \textbf{7}, 10 (2004),
  \eprint{gr-qc/0407042}.

\bibitem[Williams(2008)]{Williams:2008}
C.~Williams, \emph{Annales Henri Poincar\'e} \textbf{9}, 1029--1067 (2008).

\bibitem[Clarke(1993)]{Clarke:1993}
C.~J.~S. Clarke, \emph{The analysis of space-time singularities}, Cambridge
  University Press, 1993.

\bibitem[Hawking and Penrose(1970)]{HP1970}
S.~W. Hawking, and R.~Penrose, \emph{Proceedings of the Royal Society of
  London} \textbf{A314}, 529--548 (1970).

\bibitem[Penrose(1999)]{Penrose:1999}
R.~Penrose, \emph{Journal of Astrophysics and Astronomy} \textbf{20}, 233
  (1999).

\bibitem[Ringstr\"om(2008)]{Ringstrom:2008}
H.~Ringstr\"om, \emph{Class. Quant. Grav.} \textbf{25}, 114010 (2008).

\bibitem[Bondi et~al.(1962)]{Bondi:1962}
H.~Bondi, M.~G.~J. van~der Burg, and A.~W.~K. Metzner, \emph{Proc. Roy. Soc.
  London} \textbf{A269}, 21--52 (1962).

\bibitem[Sachs(1962)]{Sachs:1962}
R.~K. Sachs, \emph{Proc. Roy. Soc. London} \textbf{A270}, 103--26 (1962).

\bibitem[Penrose(1964)]{Penrose:1964}
R.~Penrose, \enquote{Conformal approach to infinity,} in \emph{Relativity,
  groups and topology: the 1963 les Houches lectures}, edited by B.~S. DeWitt,
  and C.~M. Dewitt, Gordon and Breach, 1964.

\bibitem[Hollands and Wald(2004)]{Hollands:2004ac}
S.~Hollands, and R.~M. Wald, \emph{Class. Quant. Grav.} \textbf{21}, 5139--5146
  (2004), \eprint{gr-qc/0407014}.

\bibitem[Penrose(1965)]{Penrose:1965}
R.~Penrose, \emph{Proc. Roy. Soc. London} \textbf{A284}, 159--203 (1965).

\bibitem[Geroch(1971)]{Geroch:1971}
R.~P. Geroch, \enquote{Space-time structure from a global point of view,} in
  \emph{General relativity and cosmology}, edited by R.~K. Sachs, Academic
  Press, 1971, vol. XLVII of \emph{Proc. Int. Sch. Physics `E. Fermi'}, pp.
  71--103.

\bibitem[Helfer(2007)]{Helfer:2007}
A.~D. Helfer, \emph{Gen. Rel. Grav.} \textbf{39}, 2125--2147 (2007),
  \eprint{arxiv:0709.1078}.

\bibitem[Penrose(1980)]{Penrose:1980}
R.~Penrose, \enquote{Singularities and time-asymmetry,} in \emph{General
  relativity: An Einstein centenary survey}, edited by S.~W. Hawking, and
  W.~Israel, Cambridge University Press, 1980.

\bibitem[Chrus\'ciel et~al.(2001)]{Chrusciel:2001}
P.~T. Chrus\'ciel, E.~Delay, G.~J. Galloway, and R.~Howard, \emph{Annales Henri
  Poincar\'e} \textbf{2}, 109--178 (2001).

\bibitem[Chrus\'ciel et~al.(2002)]{Chrusciel:2002}
P.~T. Chrus\'ciel, J.~H.~G. Fu, G.~J. Galloway, and R.~Howard, \emph{J. Geom.
  Phys.} \textbf{41}, 1--12 (2002).

\bibitem[Chrus\'ciel and Galloway(1998)]{Chrusciel:1998}
P.~T. Chrus\'ciel, and G.~J. Galloway, \emph{Communs. Math. Phys.}
  \textbf{193}, 449--470 (1998).

\bibitem[Beem and Kr\'olak(1998)]{Beem:1998}
J.~K. Beem, and A.~Kr\'olak, \emph{J. Math. Phys.} \textbf{39}, 6001--6010
  (1998).

\bibitem[Hawking(1972)]{Hawking:1971vc}
S.~W. Hawking, \emph{Commun. Math. Phys.} \textbf{25}, 152--166 (1972).

\bibitem[Chrus\'ciel et~al.(2010)]{Chrusciel:2010fn}
P.~T. Chrus\'ciel, G.~J. Galloway, and D.~Pollack, {Mathematical general
  relativity: a sampler} (2010), \eprint{arxiv:1004.1016}.

\bibitem[Carter(1970)]{Carter:1970ea}
B.~Carter, \emph{Commun. Math. Phys.} \textbf{17}, 233--238 (1970).

\bibitem[Bardeen et~al.(1973)]{Bardeen:1973gs}
J.~M. Bardeen, B.~Carter, and S.~W. Hawking, \emph{Commun. Math. Phys.}
  \textbf{31}, 161--170 (1973).

\bibitem[Wald(1984)]{Wald:1984}
R.~M. Wald, \emph{General relativity}, Chicago University Press, 1984.

\bibitem[Dunkel et~al.(2009)]{DHH:2009}
J.~Dunkel, P.~H\"anggi, and S.~Hilbert, \emph{Nature Physics} \textbf{5},
  741--747 (2009).

\bibitem[Wald(1994)]{Wald:1995yp}
R.~M. Wald, \emph{Quantum field theory in curved space-time and black hole
  thermodynamics}, University of Chicago Press, 1994.

\bibitem[Gao and Wald(2001)]{Gao:2001ut}
S.~Gao, and R.~M. Wald, \emph{Phys. Rev.} \textbf{D64}, 084020 (2001),
  \eprint{gr-qc/0106071}.

\bibitem[Helfer(2001)]{Helfer:2001}
A.~D. Helfer, \emph{Class. Quant. Grav.} \textbf{18}, 5413--5428 (2001),
  \eprint{gr-qc/0110108}.

\bibitem[Schwinger(1951)]{Schwinger:1951nm}
J.~S. Schwinger, \emph{Phys. Rev.} \textbf{82}, 664--679 (1951).

\bibitem[Unruh(1976)]{Unruh:1976db}
W.~G. Unruh, \emph{Phys. Rev.} \textbf{D14}, 870 (1976).

\bibitem[Bisognano and Wichmann(1976)]{Bisognano:1976za}
J.~J. Bisognano, and E.~H. Wichmann, \emph{J. Math. Phys.} \textbf{17},
  303--321 (1976).

\bibitem[Hawking(1974)]{Hawking:1974rv}
S.~W. Hawking, \emph{Nature} \textbf{248}, 30--31 (1974).

\bibitem[Hawking(1975)]{Hawking:1974sw}
S.~W. Hawking, \emph{Commun. Math. Phys.} \textbf{43}, 199--220 (1975).

\bibitem[Helfer(2003)]{Helfer:2003va}
A.~D. Helfer, \emph{Rept. Prog. Phys.} \textbf{66}, 943--1008 (2003),
  \eprint{gr-qc/0304042}.

\bibitem[Wald(1975)]{Wald:1975kc}
R.~M. Wald, \emph{Commun. Math. Phys.} \textbf{45}, 9--34 (1975).

\bibitem[Bekenstein and Mukhanov(1995)]{Bekenstein:1995ju}
J.~D. Bekenstein, and V.~F. Mukhanov, \emph{Phys. Lett.} \textbf{B360}, 7--12
  (1995), \eprint{gr-qc/9505012}.

\bibitem[Ashtekar(1998)]{Ashtekar:1998}
A.~Ashtekar, \enquote{Geometric issues in quantum gravity,} in \emph{The
  geometric universe: science, geometry and the work of Roger Penrose}, edited
  by S.~A. Huggett, L.~J. Mason, K.~P. Tod, S.~T. Tsou, and N.~M.~J. Woodhouse,
  Oxford University Press, 1998.

\bibitem[Hawking(1976)]{Hawking:1976de}
S.~W. Hawking, \emph{Phys. Rev.} \textbf{D13}, 191--197 (1976).

\bibitem['t~Hooft(1996)]{Hooft:1996tq}
G.~'t~Hooft, \emph{Int. J. Mod. Phys.} \textbf{A11}, 4623--4688 (1996),
  \eprint{gr-qc/9607022}.

\bibitem[Hawking and Penrose(1996)]{Hawking:1996jh}
S.~Hawking, and R.~Penrose, \emph{The nature of space and time}, Princeton
  University Press, 1996.

\bibitem[Parikh and Wilczek(2000)]{Parikh:1999mf}
M.~K. Parikh, and F.~Wilczek, \emph{Phys. Rev. Lett.} \textbf{85}, 5042--5045
  (2000), \eprint{hep-th/9907001}.

\bibitem[Helfer(2004{\natexlab{a}})]{Helfer:2004jx}
A.~D. Helfer, \emph{Phys. Lett.} \textbf{A329}, 277--283 (2004{\natexlab{a}}),
  \eprint{gr-qc/0407055}.

\bibitem[MacGibbon and Webber(1990)]{MacGibbon:1990zk}
J.~H. MacGibbon, and B.~R. Webber, \emph{Phys. Rev.} \textbf{D41}, 3052--3079
  (1990).

\bibitem[Gibbons and Perry(1976)]{Gibbons:1976es}
G.~W. Gibbons, and M.~J. Perry, \emph{Phys. Rev. Lett.} \textbf{36}, 985
  (1976).

\bibitem[Helfer(2004{\natexlab{b}})]{Helfer:2005wy}
A.~D. Helfer, \emph{Int. J. Mod. Phys.} \textbf{D13}, 2299--2305
  (2004{\natexlab{b}}), \eprint{gr-qc/0503052}.

\bibitem[Helfer(2005)]{Helfer:2005wz}
A.~D. Helfer  (2005), \eprint{gr-qc/0503053}.

\bibitem[Helfer(1996)]{Helfer:1996my}
A.~D. Helfer, \emph{Class. Quant. Grav.} \textbf{13}, L129--L134 (1996),
  \eprint{gr-qc/9602060}.

\bibitem[Helfer(1998)]{Helfer:1998}
A.~D. Helfer, \emph{Class. Quant. Grav.} \textbf{15}, 1169--83 (1998).

\bibitem[Lamoreaux(2010)]{Lamoreaux:2010}
S.~K. Lamoreaux, Progress in experimental measurements of the surface-surface
  casimir force: Electrostatic calibrations and limitations to accuracy (2010),
  \eprint{arxiv:1008.3640}.

\bibitem[Klimchitskaya and Mostopanenko(2010)]{KM:2010}
G.~L. Klimchitskaya, and V.~M. Mostopanenko (2010),
  \eprint{arxiv:1010.2216v1}.

\end{thebibliography}

\end{document}